\documentclass[showpacs,preprint]{revtex4}
\usepackage{graphicx}
\usepackage{dcolumn}
\usepackage{amsmath}
\usepackage[latin1]{inputenc}
\usepackage{graphicx, psfrag}
\usepackage{amssymb}
\usepackage[colorlinks=true, citecolor=blue, urlcolor = blue, linkcolor= red, bookmarks=true]{hyperref}
\usepackage{float}
\usepackage{amsmath}
\usepackage{amsfonts}
\usepackage{dcolumn}
\usepackage{hyperref}
\usepackage{subfigure}
\usepackage{pgfplots}
\usepackage{epstopdf}
\usepackage{booktabs}

\makeatletter
\def\btt#1{\texttt{\@backslashchar#1}}
\DeclareRobustCommand\bblash{\btt{\@backslashchar}} \makeatother

\begin{document}

\title[]{Shadow cast and deflection of light by charged rotating regular black holes}
\author{Rahul Kumar$^{a}$}\email{rahul.phy3@gmail.com}

\author{Sushant~G.~Ghosh$^{a,\;b}$} \email{sghosh2@jmi.ac.in, sgghosh@gmail.com}
\author{Anzhong Wang${}^{c, d}$}\email{anzhong$\_$wang@baylor.edu}

\affiliation{$^{a}$ Centre for Theoretical Physics, Jamia Millia
	Islamia, New Delhi 110025, India}
\affiliation{$^{b}$ Astrophysics and Cosmology
	Research Unit, School of Mathematics, Statistics and Computer Science, University of
	KwaZulu-Natal, Private Bag 54001, Durban 4000, South Africa}
\affiliation{ ${}^{c}$ GCAP-CASPER, Physics Department, Baylor University, Waco, Texas 76798-7316, USA\\
	${}^{d}$ Institute for Theoretical Physics and Cosmology, Zhejiang University of Technology, Hangzhou, 310032, China}

\date{\today}

\begin{abstract}
We discuss the horizon properties, shadow cast, and the weak gravitational lensing of charged rotating regular black holes, which in addition to mass ($M$) and rotation parameter ($a$) have an electric charge ($Q$) and magnetic charge ($g$). The considered regular black holes are the generalization of the Kerr ($Q=g=0$) and Kerr-Newman ($g=0$) black holes. Interestingly, for a given parameter set, the apparent size of the shadow monotonically decreases and the shadow gets more distorted with increasing charge parameter $Q$. We put constraints on the black hole parameters with the aid of recent M87* shadow observation. The conserved quantities associated with the rotating regular black holes are calculated and also a brief description of the weak gravitational lensing using the Gauss-Bonnet theorem is presented. Interestingly, the deflection angle decreases with the charge of the black hole.  Our results \textit{vis-\`{a}-vis} go over to the Kerr and Kerr-Newman black holes in the appropriate limits. 
\end{abstract}

\maketitle
\section{Introduction}
The celebrated singularity theorem asserts that the gravitational collapse of the sufficiently massive stars, under certain conditions, necessarily leads to the formation of spacetime singularities \cite{Hawking:1969sw,Penrose:1964wq,Hawking:1967ju}. One of the pathologies in the classical general relativity is the inevitable existence of singularities. The well-known black hole solutions of general relativity such as the Schwarzschild, Reissner-Nordstr\"om, and Kerr metrics harbor curvature singularities in the interior. It leads to the belief that the classical general relativity requires modifications where spacetime curvature blows, and that these singularities are likely to be resolved by quantum gravity. In the absence of a well-defined satisfactory theory of quantum gravity, to understand and resolve the internal singularity of a  black hole,  attentions were shifted to regular models, which are motivated by quantum arguments. Sakharov \cite{Sakharov:1966aja} and Gliner \cite{Gliner} proposed that for small $r$, the Einstein tensor is $G_{\mu \nu} = \Lambda g_{\mu \nu}$, with $\Lambda \neq 0$, or equivalently the requirement that there exists a central de Sitter core obeying the equation of state $P=-\rho$, and eventually putting an upper bound on the scalar curvature. Consequently, the final collapsed state, a metastable state balancing the de Sitter outward radial pressure and the gravitational inward pressure, exterminates such curvature singularity at its center, and in the presence of horizon imitates a regular black hole.  Bardeen \cite{Bardeen1} realized this idea to propose the first-ever model for the regular black hole, which is later shown to be an exact solution of Einstein's field equations coupled with the nonlinear electrodynamics (NED) \cite{AyonBeato:1998ub,AyonBeato:1999ec,AyonBeato:2000zs}. Since then several other regular black hole solutions have been proposed and discussed  \cite{Dymnikova:1992ux,Dymnikova:2004zc,Bronnikov:2000vy,Bronnikov:2005gm,Berej:2006cc,Burinskii:2002pz} (see also \cite{Ansoldi:2008jw} for a general review of regular black holes). It turns out that the absence of central singularities makes the Hawking radiation and information loss issues less problematic \cite{Ashtekar:2005cj}.  Later, Hayward \cite{Hayward:2005gi} proposed a concrete model, which is convenient for the analysis, of the collapse and evaporation phases. The formation of a  black
hole from an initial vacuum region is described by a  regular black hole, which can be also obtained by mainly working
within NED. 

In general an action minimally coupled with NED \cite{AyonBeato:2000zs} is given by
\begin{equation}
I=\frac{1}{16\pi}\int d^4 x\sqrt{-g}\left(R-\mathcal{L(F)}\right),\label{actionEq}
\end{equation}
where $R$ is the Ricci curvature scalar, and the Lagrangian density $\mathcal{L(F)}$ is a function of $\mathcal{F}=F^{\mu\nu}F_{\mu\nu}/4$ with $F_{\mu\nu}=\partial_{\mu}A_{\nu}-\partial_{\nu}A_{\mu}$ being the electromagnetic field tensor for the gauge potential $A_{\mu}$. On varying action (\ref{actionEq}), the field equations of motion read \cite{AyonBeato:2000zs, AyonBeato:1999ec}
\begin{eqnarray}
G_{\mu\nu}&=&T_{\mu\nu}=2\left(\mathcal{L_F}F_{\mu\sigma}F_{\nu}^{\sigma}-\frac{1}{4}g_{\mu\nu}\mathcal{L(F)}\right)~\label{Eq1}\\
&&	\nabla_{\mu}\left(\mathcal{L_F}F^{\mu\nu}\right)=0 \quad \text{and}\quad \nabla_{\mu}(^*F^{\mu\nu})=0.\label{Eq2}
\end{eqnarray}
We assume the static and spherically symmetric metric \textit{anstaz} as
\begin{equation}
ds^2=-\left(1-\frac{2m(r)}{r}\right)dt^2+\left(1-\frac{2m(r)}{r}\right)^{-1}dr^2+r^2(d\theta^2+\sin^2\theta d\phi^2).\label{Eq3}
\end{equation}
To derive the Hayward black hole \cite{Hayward:2005gi}, we choose the Lagrangian density \cite{Fan:2016hvf}
\begin{equation}
\mathcal{L(F)}=\frac{6}{sg^2}\frac{(2g^2\mathcal{F})^{3/2}}{\left(1+(2g^2\mathcal{F})^{3/4}\right)^2},\label{Eq001}
\end{equation}
where $s$ is a constant and $g$ is the magnetic charge. The Maxwell field tensor is
\begin{equation}
F_{\mu\nu}=2\delta^{\theta}_{[\mu}\delta^{\phi}_{\nu]}g(r)\sin\theta.
\end{equation}
With this choice of $F_{\mu\nu}$, the Maxwell Eq.~(\ref{Eq2}) implies
\begin{equation} 
g'(r)\sin\theta dr\wedge d\theta\wedge d\phi=0,\label{Eqcharge}
\end{equation}
where $\prime$ corresponds to the derivative with respect to $r$. Equation (\ref{Eqcharge}) leads to $g(r)=g=$constant. Then the field tensor and $\mathcal{F}$ become
\begin{equation}
F_{\theta\phi}=g\sin\theta,\quad \mathcal{F}=\frac{g^2}{2r^4}, \label{Eq002}
\end{equation} 
where the magnetic charge $g$ is defined as $\int \mathcal{F}/{4\pi}=g$. Now, with Eqs.~(\ref{Eq001}) and (\ref{Eq002}), one gets
\begin{equation}
\mathcal{L}=\frac{6 g^4}{s(r^3+g^3)^2}\label{Lag}
\end{equation}
On using Eqs.~(\ref{Eq1}) and (\ref{Lag}), the energy-momentum tensor reads
\begin{eqnarray}
T_{t}^{t}=T_{r}^{r}=-\frac{3 g^4}{s(r^3+g^3)^2},\nonumber\\
T_{\theta}^{\theta}=T_{\phi}^{\phi}=-\frac{3 g^4(-11r^3+g^3)}{s(r^3+g^3)^3}.\label{EMT}
\end{eqnarray}  
Using the metric \textit{anstaz} (\ref{Eq3}), Eq.~(\ref{Eq1}) with energy-momentum tensor Eq.~(\ref{EMT}) admits an exact solution
\begin{equation}
ds^2=-\left(1-\frac{2Mr^2}{r^3+2M\ell^2}\right)dt^2+\left(1-\frac{2Mr^2}{r^3+2M\ell^2}\right)^{-1}dr^2+r^2(d\theta^2+\sin^2\theta d\phi^2),\label{Eq004}
\end{equation}
which is the Hayward black hole \cite{Hayward:2005gi}. Here, $s=|g|/2M$, $g^3=2M \ell^2$ and $M$ appears as an integration constant and can be identified as the black hole mass parameter \cite{Hayward:2005gi,Fan:2016rih}. The magnetic charge $g$ is indeed related to the length associated with the region concentrating the central energy density, such that modifications in the spacetime metric appear when the curvature scalar becomes comparable with $\ell^{-2}$ \cite{Hayward:2005gi}. Moreover, the nonzero value of $\ell$ prevents the curvature scalars to grow infinitely at the central region and makes them bounded from above, just in the spirit of the original idea of the regular black hole.

Recently, Frolov \cite{Frolov:2016pav} included the electric Maxwell charge, and proposed the charged Hayward black hole, which reads 
\begin{equation}
ds^2=-\left(1-\frac{\left(2Mr-Q^2\right)r^2}{r^4+\left(2Mr+Q^2\right)\ell^2}\right)dt^2+\left(1-\frac{\left(2Mr-Q^2\right)r^2}{r^4+\left(2Mr+Q^2\right)\ell^2}\right)^{-1}dr^2+r^2(d\theta^2+\sin^2\theta d\phi^2).\label{CHmetric}
\end{equation}
In the limit $\ell\to 0$, one recovers the Reissner-Nordstr\"{o}m metric. Further, at $r\to \infty$ and $r\to 0$, one has, respectively,
\begin{eqnarray}
g_{tt}&=&1-\frac{2M}{r}+\frac{Q^2}{r^2} +\ell^2\mathcal{O}(r^{-4}),\\
g_{tt}&=&1+\frac{r^2}{\ell^2}+\mathcal{O}(r^6),
\end{eqnarray}
which implies that the charged Hayward metric is also regular at the origin with curvature being of the order $\ell^{-2}$. The causal structure of the Hayward black hole makes resemblance with that of the Reissner-Nordstr\"{o}m spacetime, except that now $r=0$ is not singular anymore \cite{DeLorenzo:2014pta}. Subsequently, the Hayward black hole has been studied in the wide context of physical phenomenon \cite{Toshmatov:2015wga,Lin:2013ofa,DeLorenzo:2014pta,Neves:2019ywx,Contreras:2018gpl,Frolov:2016pav}. The generalization of these static Hayward black holes to the axially symmetric case, Kerr-like black hole, was also addressed recently \cite{Bambi:2013ufa, Amir:2015pja, Ghosh:2014hea}.  It is demonstrated \cite{Bambi:2013ufa} that
the rotating Hayward black holes can be derived
starting from exact spherical solutions (\ref{Eq004}) by a
complex coordinate transformation due to 
Newman and Janis \cite{Newman:1965tw,Azreg-Ainou:2014pra} in general relativity. Gravitational lensing and a black hole shadow provide possible ways to distinguish the regular black holes from the Kerr black hole \cite{ Abdujabbarov:2016hnw,Lamy:2018zvj,Jusufi:2018jof,Ovgun:2019wej,Li:2013jra,Tsukamoto:2014tja,Kumar:2018ple}.  

The supermassive black holes at the galactic centers are believed to possess a finite electric charge, which can have considerable effects on the electromagnetic processes around its vicinity \cite{Zakharov:2014lqa,Zakharov:2018syb,Zajacek:2018ycb}. In this paper, we obtain and discuss the charged rotating Hayward black holes, with an implication to test them with astrophysical observations. It is natural to anticipate the rich spectrum of black hole physics with the inclusion of electric charge and rotation parameter to the regular black holes and to check if a charged one can be observationally distinguished from the uncharged counterpart. We discuss the horizon structures and investigate the issue of energy conditions. The study of null geodesics around black holes is of great importance both from the theoretical and observational point of view, as they play crucial roles in determining the strong gravitational field features, such as gravitational lensing and shadow. We calculate the deflection angle of light by considering the source and observer at finite distances from the black hole and compare it to that for the Kerr-Newman and Kerr black holes. With the aid of recent shadow observational data from the EHT Collaboration \cite{Akiyama:2019cqa,Akiyama:2019eap}, we modeled the charged rotating Hayward black hole as M87* and constrained the parameter space which at best can describe the observed asymmetry of the shadow. We find that for a certain parameter space the charged rotating Hayward black hole resembles the observed image. 

The paper is organized as follows. The Horizon structures and the issue of energy condition violation are investigated in Sec.~\ref{sec3}. Section~\ref{sec4} is devoted to the discussion of the Komar conserved quantities. In Secs.~\ref{sec5} and \ref{sec6}, we discuss the possible observational consequences specifically the black hole shadow and the gravitational lensing phenomenon around a charged rotating Hayward black hole. Finally, we conclude Sec.~\ref{sec7}.  In this paper, we used the geometric system of units $G=\hbar=c=1$.

\section{Horizon properties and weak energy condition}\label{sec3}

 The metric (\ref{CHmetric}) allows a generalization to the stationary and axially symmetric spacetime, namely, the charged rotating Hayward black hole, which in the Boyer-Lindquist coordinates reads  \cite{Bambi:2013ufa,Toshmatov:2014nya,Toshmatov:2017zpr} 
\begin{eqnarray}\label{rotbhtr}
ds^2 & = & - \left( 1- \frac{2m(r)r}{\Sigma} \right) dt^2  - \frac{4am(r)r}{\Sigma  } \sin^2 \theta dt \; d\phi +
\frac{\Sigma}{\Delta}dr^2 + \Sigma d \theta^2 \nonumber
\\ && +  \left(r^2+ a^2 +
\frac{2m(r) r a^2 }{\Sigma} \sin^2 \theta
\right)\sin^2 \theta d\phi^2,
\end{eqnarray}
where $\Sigma = r^2 + a^2 \cos^2\theta$, $\Delta=r^2 + a^2 - 2m(r)r
$ and $a$ is the spin parameter. The metric has the form of Kerr black hole with mass $m(r)$, which is a measure of mass inside the region of constant radial coordinate $r$, such that in the limit $r\to\infty$, it approaches the black hole mass parameter $M$, i.e. $\lim_{r\to \infty}m(r)=M$. Indeed, the mass function interpolates between a de-Sitter core and the asymptotically flat spatial infinity \cite{Frolov:2016pav}. The charged rotating Hayward black hole metric is a prototype of a large class of the Kerr family, where $\ell$ describes the deviation from the Kerr-Newman black hole \cite{Newman:1965my}, which is recovered when $\ell=0$. The Kerr black hole solution \cite{kerr} can be realized from metric (\ref{rotbhtr}), as the special case when $\ell=Q=0$. The Reissner-Nordstr\"{o}m \cite{Reissner} and Schwarzschild solutions are special cases, respectively, when $a=\ell=0$ and $Q=a=\ell=0$. Besides, at large distance ($r>> ((2Mr+Q^2)\ell^2)^{1/4}$) the metric correctly reproduces the Kerr-Newman solution, nevertheless, at the asymptotic spatial infinity, it goes over to the Minkowski spacetime.

The black hole metric (\ref{rotbhtr}) is stationary and axisymmetric, which entails the two obvious isometries of the spacetime. The generators of these time-translational and rotational symmetries along the $t$ and $\phi$ axis, respectively, represent the Killing vectors $\xi_{(t)}^{\mu}=\delta_{t}^{\mu}$ and $\xi_{(\phi)}^{\mu}=\delta_{\phi}^{\mu}$, and one has
\begin{eqnarray}
\xi^{\mu}_{(t)}\xi_{(t)\mu}&=&g_{tt}=-\left[1-\frac{2m(r)r}{\Sigma}\right],\\
\xi^{\mu}_{(t)}\xi_{(\phi)\mu}&=&g_{t\phi}=-\frac{2m(r)ar\sin^2\theta}{\Sigma},\\
\xi^{\mu}_{(\phi)}\xi_{(\phi)\mu}&=&g_{\phi\phi}=\left[r^2+a^2+\frac{2m(r)ra^2\sin^2\theta}{\Sigma}\right]\sin^2\theta.
\end{eqnarray}
The static limit surface (SLS) or infinite redshift surface of a black hole (\ref{rotbhtr}) is a surface at which the time translational Killing vector becomes null, such that
\begin{eqnarray}
r^2 +a^2\cos^2\theta-2m(r)r=0.
\end{eqnarray}
\begin{figure}[H]
	\begin{tabular}{c c}
		\includegraphics[scale=0.65]{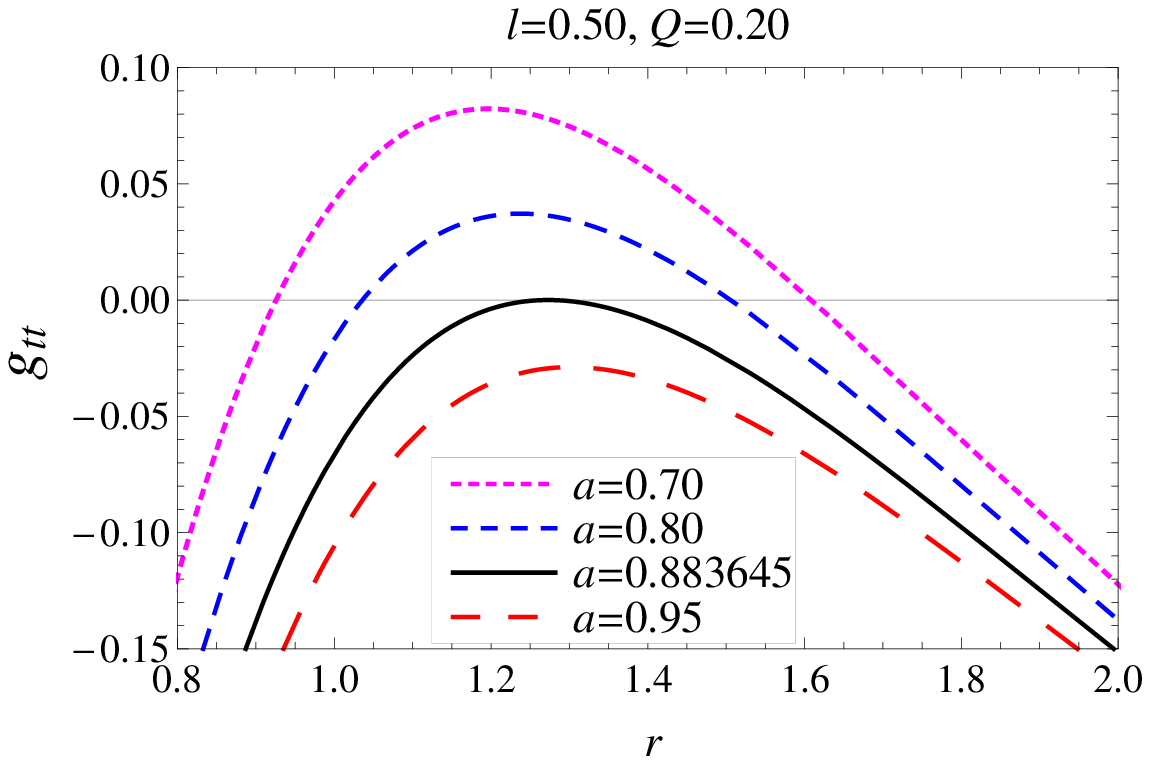}&
		\includegraphics[scale=0.65]{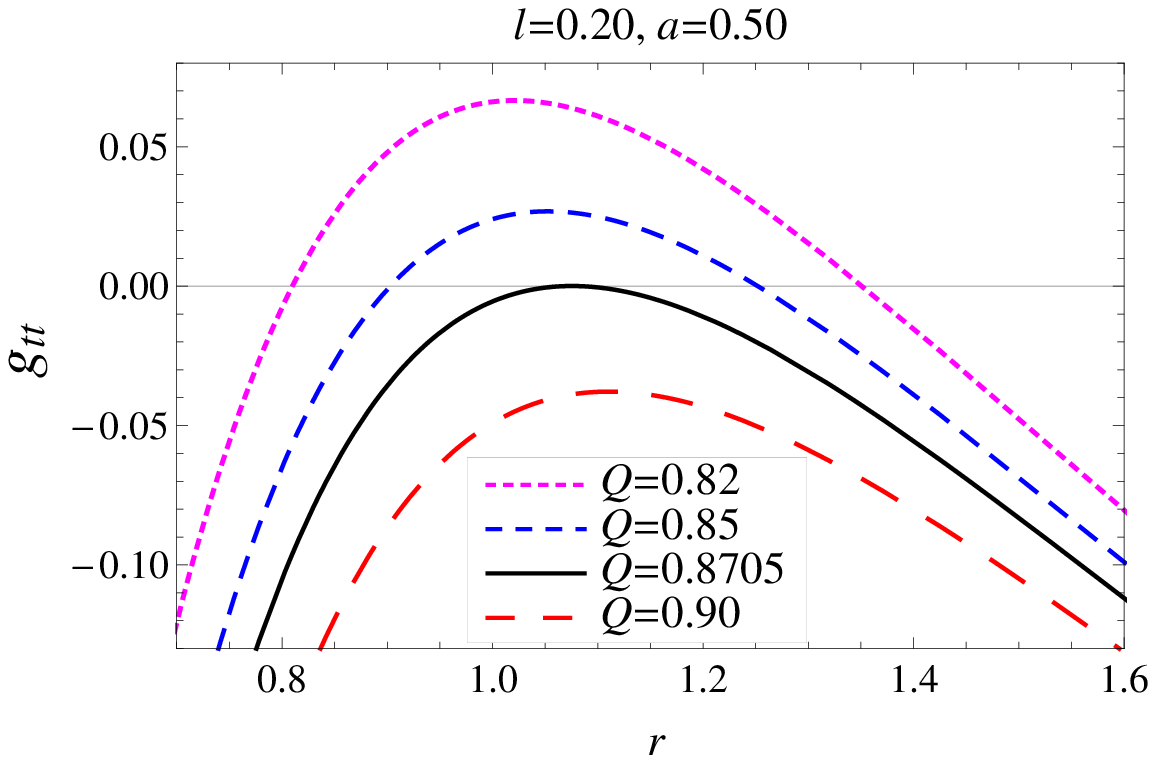}
	\end{tabular}
	\caption{The behavior of the metric function $g_{tt}$ vs $r$ for some given parameters $a, Q$, and $\ell$. The black solid curve in each plot corresponds to the degenerate SLS.} \label{gtt}
\end{figure}

Numerical analysis of $g_{tt}=0$ reveals that depending upon the values of the black hole parameters there exist three different cases of particular interests: (i) two roots $r_S{^{+}}\; \text{and} \;r_S{^{-}}$, corresponds to the outer and inner SLS of the black hole with $r_S{^{+}}> r_S{^{-}}$, (ii) degenerate SLS when $r_S{^{+}}=r_S{^{-}}$, and (iii) no SLS, when both $r_S{^{+}}\; \text{and} \;r_S{^{-}}$ become unphysical. The possible roots of $g_{tt}=0$ with different parameter combinations are depicted in Fig.~\ref{gtt}. It is shown that two SLS come closer with increasing $a$ or $Q$ and eventually coincided.

\begin{figure}
	 \begin{tabular}{c c}
\includegraphics[scale=0.65]{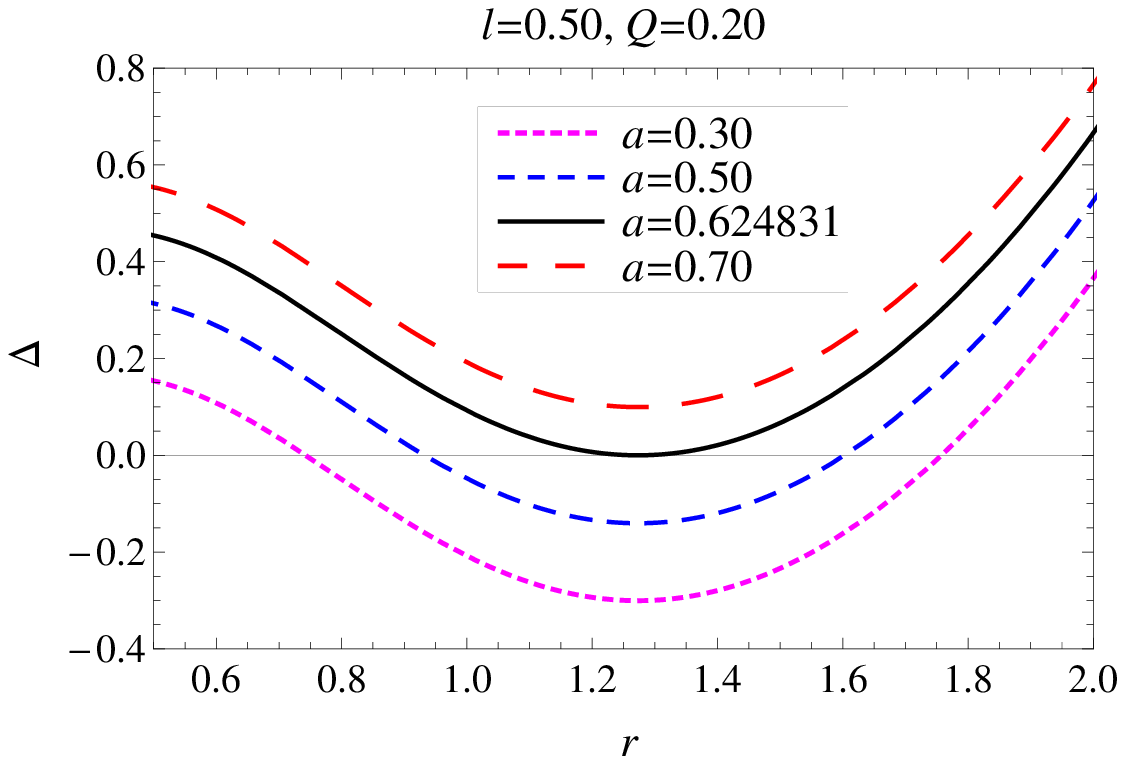}&
\includegraphics[scale=0.65]{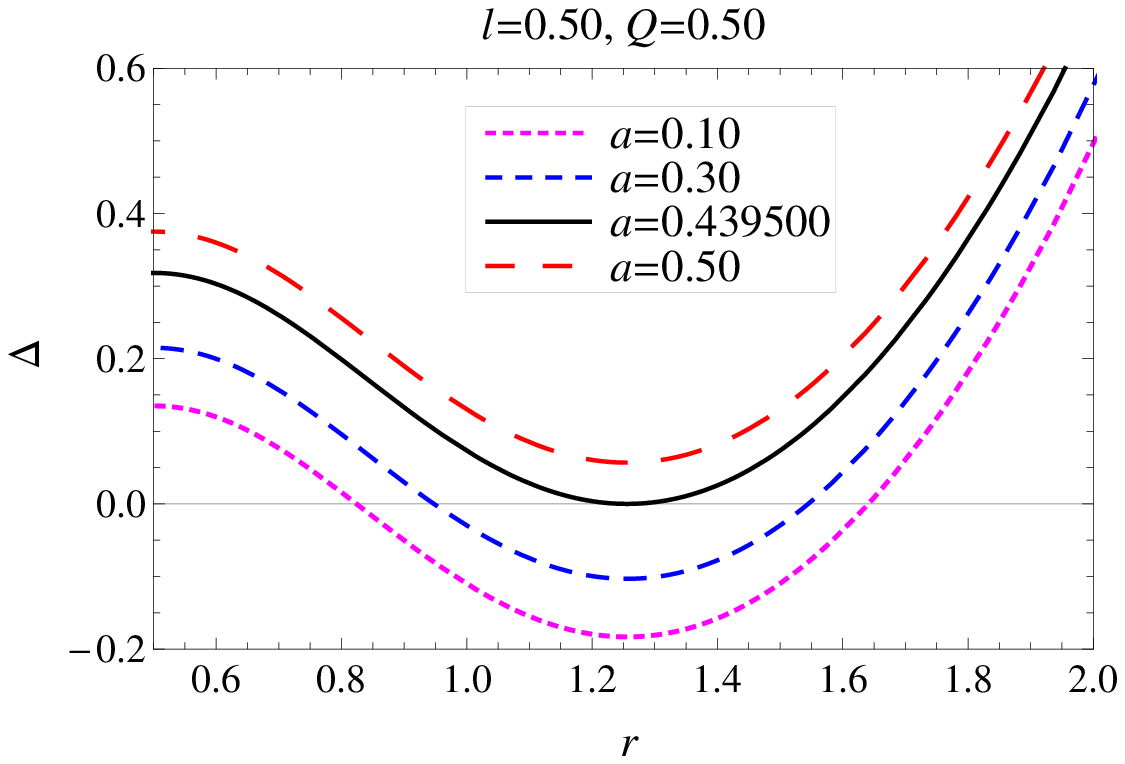}\\
\includegraphics[scale=0.65]{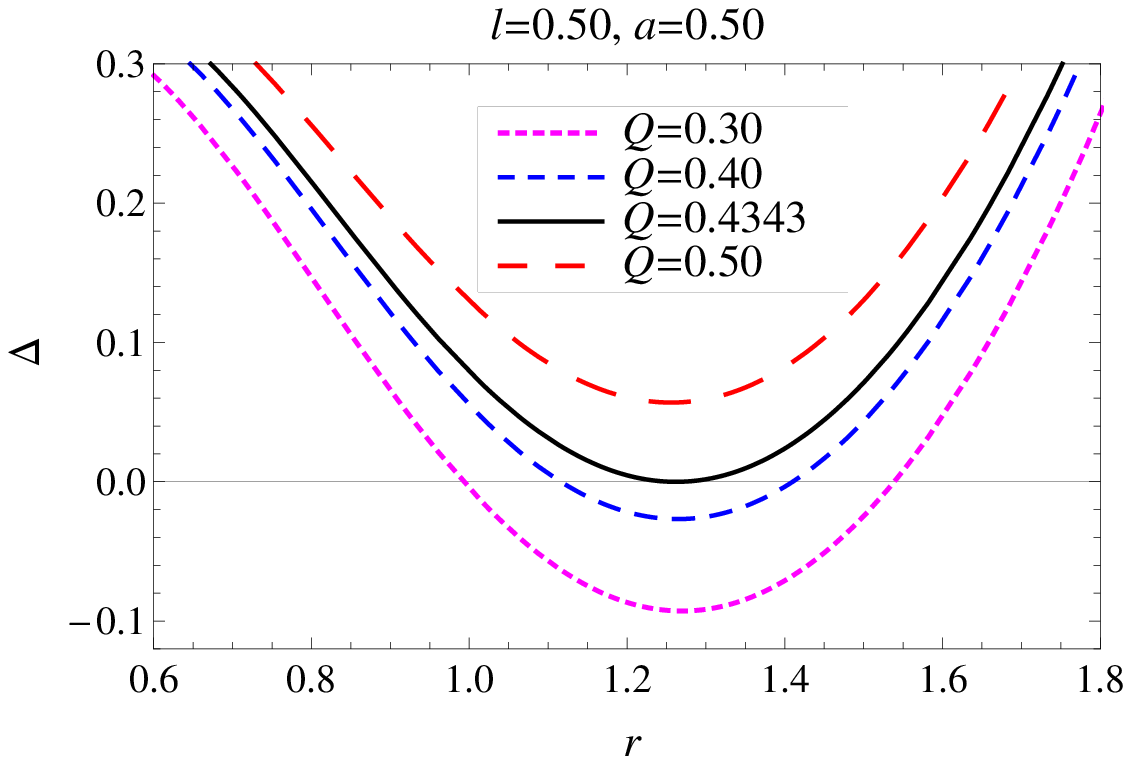}&
\includegraphics[scale=0.65]{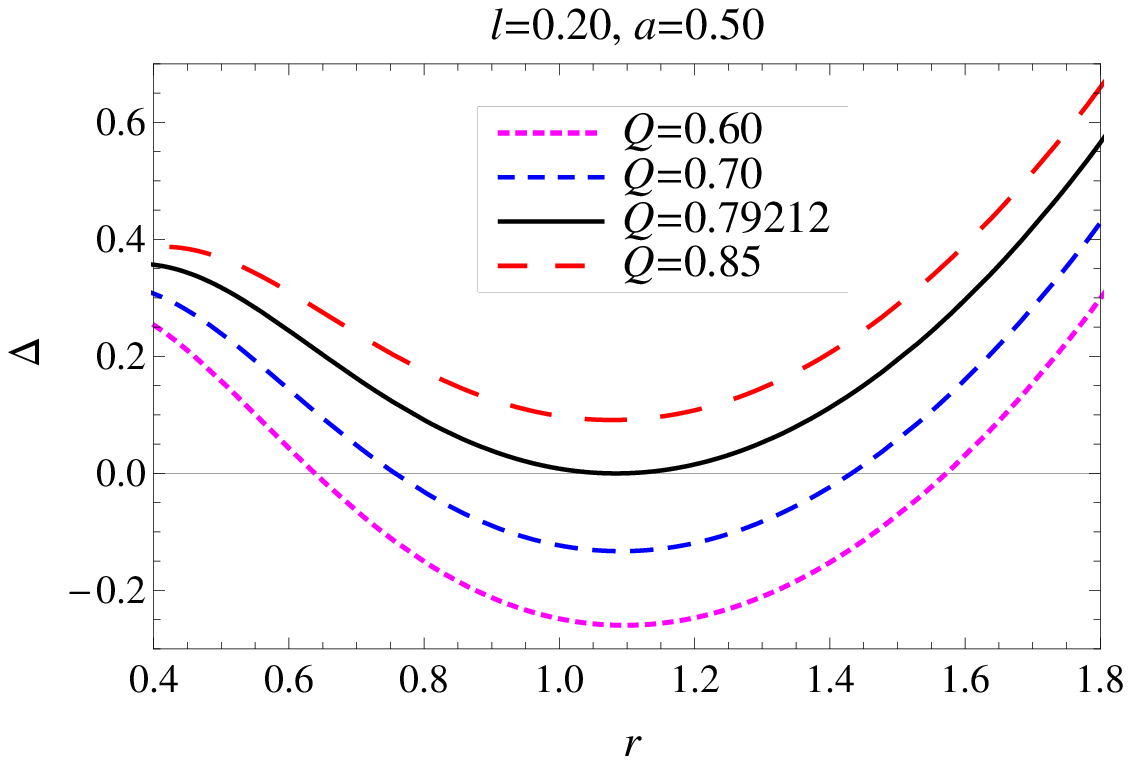}\\
\includegraphics[scale=0.65]{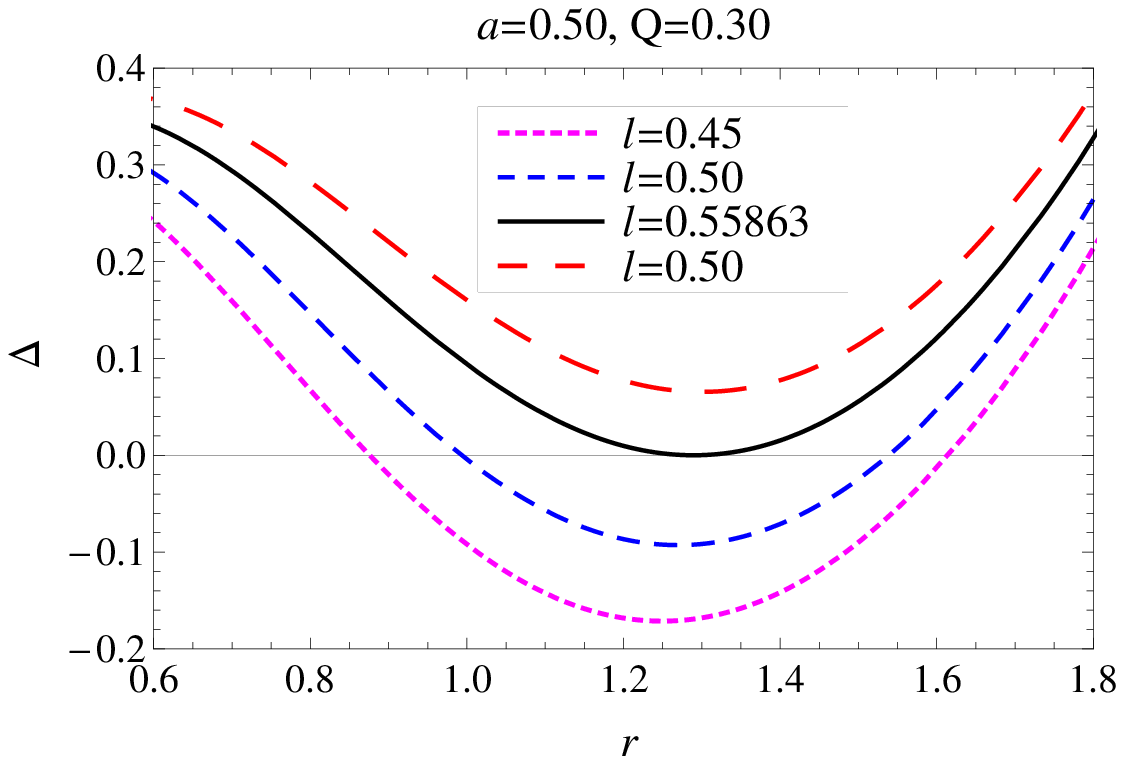}&
\includegraphics[scale=0.65]{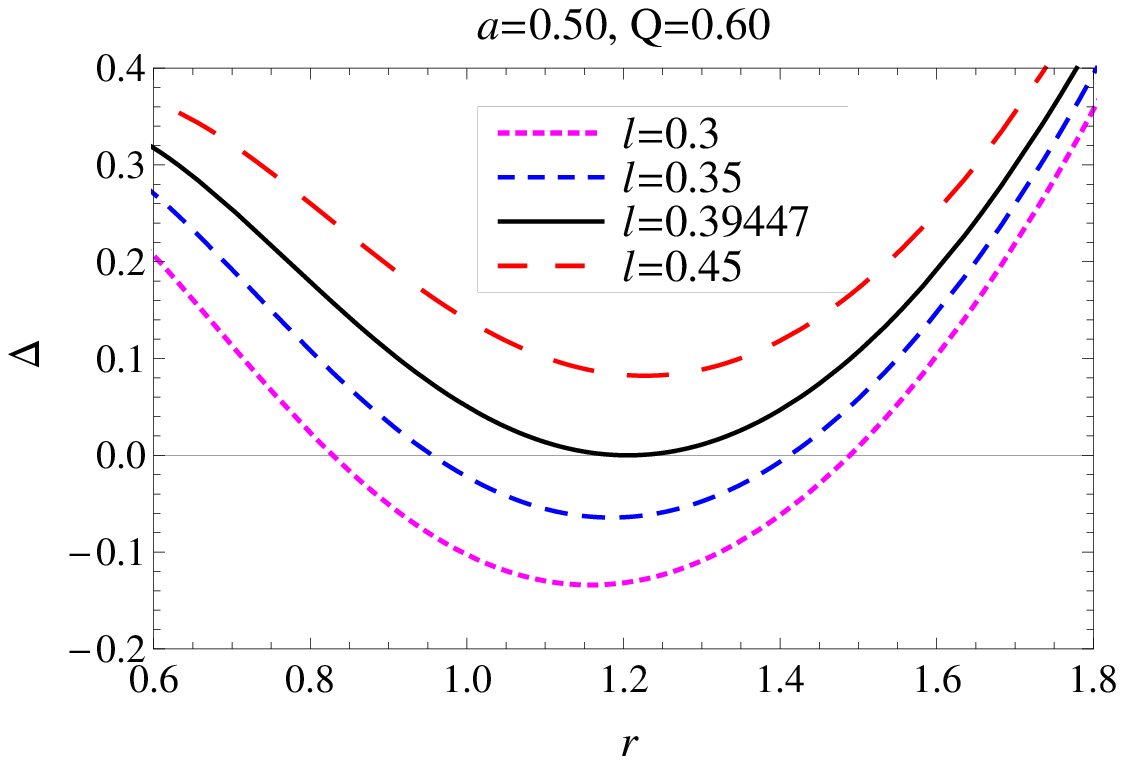}
     \end{tabular}
  \caption{The behavior of $\Delta$ with $r$ for varying parameters $a, Q$, and $\ell$. The black solid line corresponds to the extremal black hole with degenerate horizons.}
\label{Horizonfig}
\end{figure}

\begin{figure*}
	\includegraphics[scale=0.8]{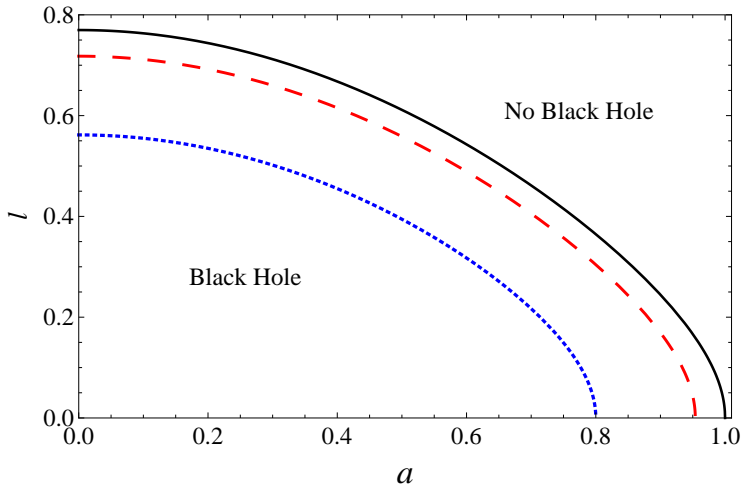}
	\caption{The parameter space of $a$ and $\ell$ for different values of $Q=0.0$ (black solid line),  $Q=0.3$ (red dashed line) and $Q=0.6$ (blue dotted line).}
	\label{NoBHfig}
\end{figure*}
The rotating metric (\ref{rotbhtr}), like the Kerr metric, is also singular at $\Delta=0$ corresponding to the position of horizons, which are determined by solving
\begin{eqnarray}
r^2 +a^2-2m(r)r=0.\label{Delta}
\end{eqnarray} 
When Eq.~(\ref{Delta}) has two roots, they correspond to the inner Cauchy horizon ($r_-$) and outer event horizon ($r_+$) and represent the nonextremal black hole, while no black hole solution exists when Eq.~(\ref{Delta}) admits no real positive roots, i.e., no horizon exists. Obviously, when Eq.~(\ref{Delta}) has a double root, the two horizons coincide and correspond to the extremal black hole. The behavior of the horizons are shown in Fig.~\ref{Horizonfig} for different parameters. Interestingly enough the extremal values of the black hole parameters for which two SLS coincide are strikingly different from those values which lead to degenerate horizons (cf. Fig.~\ref{Horizonfig}).  It is evident that the horizons can have three different possible configurations,  namely, two distinct roots for $r_-$ and $r_+$ ($a<a_E$), degenerate horizons with $r_+=r_-\equiv r_+^{E}$ ($a=a_E$), and no real roots for $r_+$ and $r_-$ ($a>a_E$).

\begin{figure*}
	\begin{tabular}{c c}
		\includegraphics[scale=0.65]{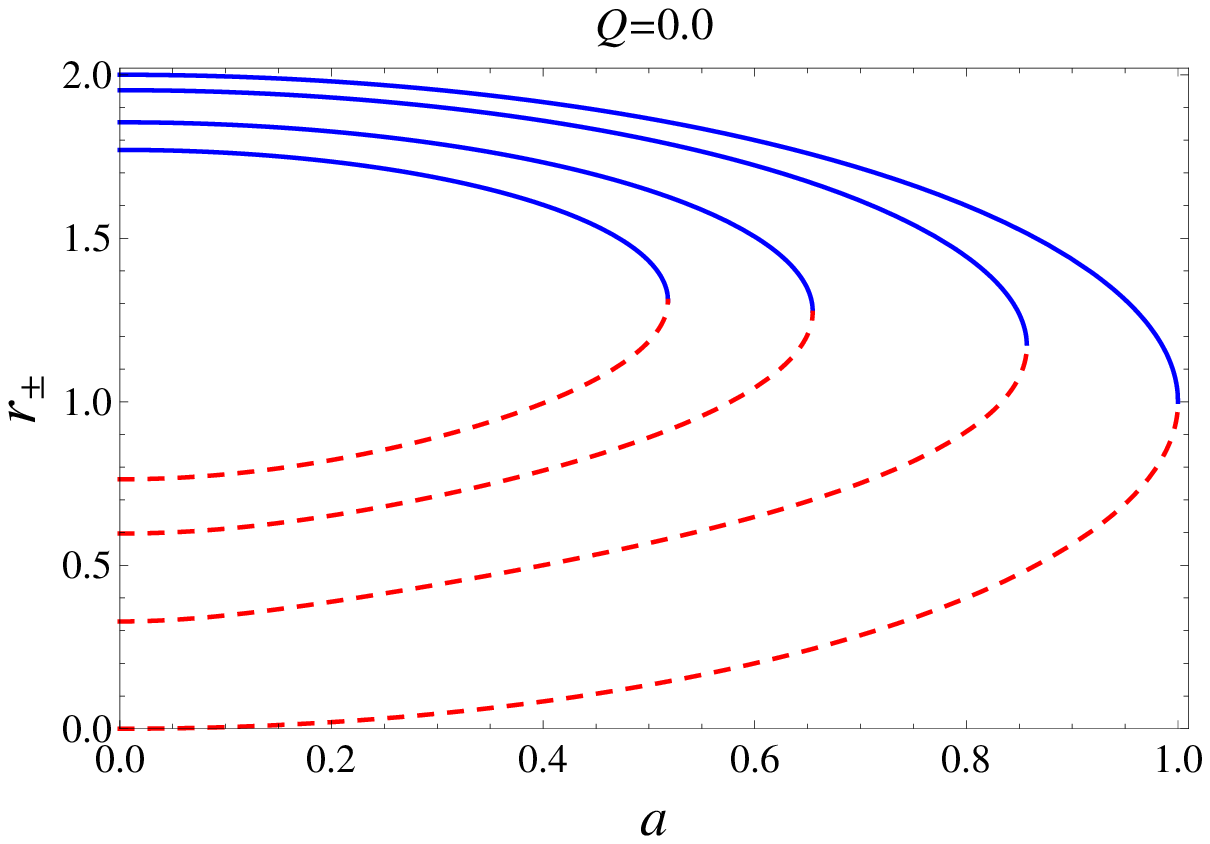}&
		\includegraphics[scale=0.65]{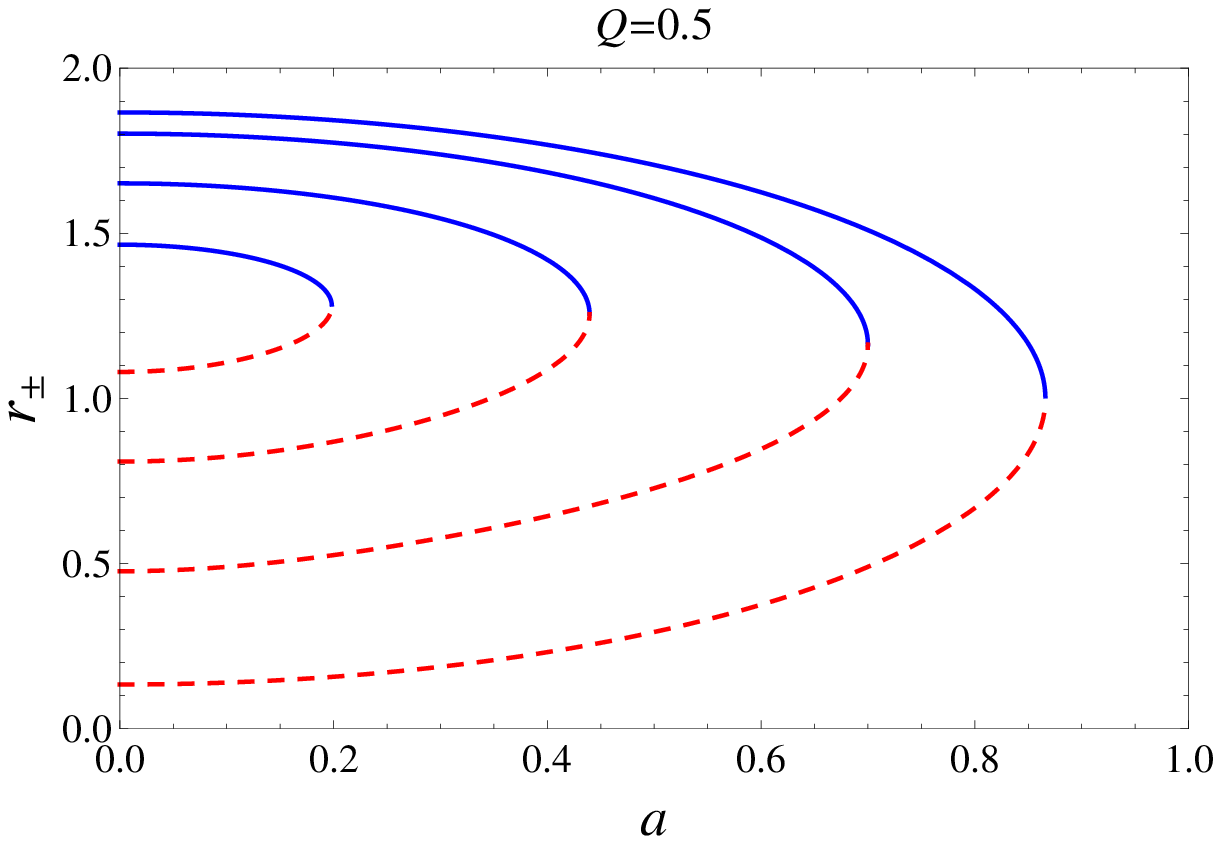}
	\end{tabular}
	\caption{The event horizon (blue solid line) and Cauchy horizon (red dashed line) radii vs  $a$ for different values of $Q$ and $\ell=0.0, 0.3, 0.5, 0.6$ from the outer to inner curves. The outermost curve represents the Kerr (left panel) and Kerr-Newman  black hole (right panel). }
\label{horizonradius}
\end{figure*}

Accordingly, we find that for fixed $Q$ the whole parameter space ($a, \ell$) can be divided into two regions as shown in Fig.~\ref{NoBHfig}. Such that, parameters inside each curve yield two distinct horizons, whereas those outside correspond to the no-horizon configurations. Each curve delineates the boundary between these two regions and comprises the extremal values of the parameters which lead to the existence of extremal black holes. Figure \ref{NoBHfig} infers that the charge $Q$ has a profound effect on the allowed range of the parameters $a$ and $\ell$ for the presence of the black hole horizons, as the allowed parameter space gradually decreases with increasing $Q$. For fixed values of the parameters, the event horizon radius of the charged rotating Hayward black hole is smaller than those of the Kerr-Newman and uncharged rotating Hayward black hole, as is evident from Fig.~\ref{horizonradius}.   

\begin{figure}
	\begin{tabular}{c c}
		\includegraphics[scale=0.72]{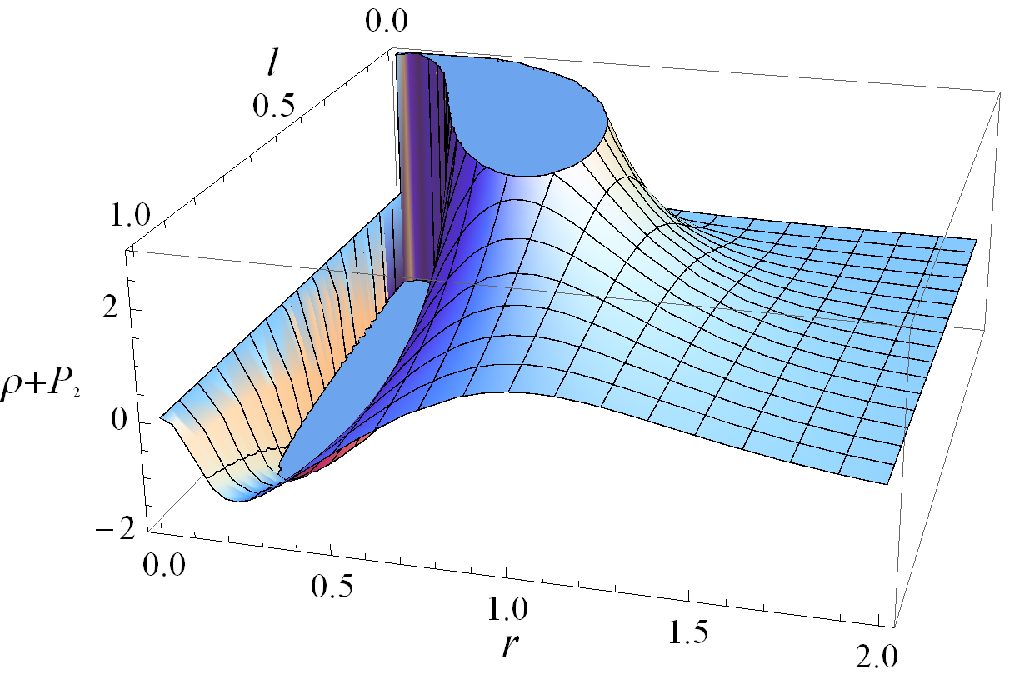}&
		\includegraphics[scale=0.72]{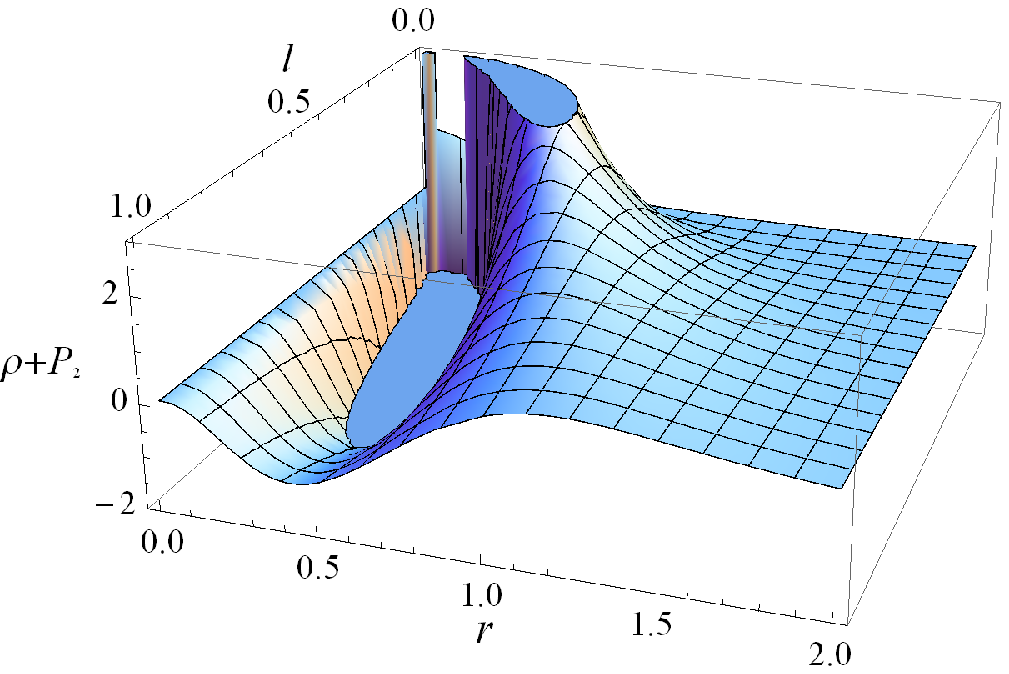}
	\end{tabular}
	\caption{The behavior of $\rho+P_2$ with $r$ and $\ell$ for $a=0.20$ and different values of $Q$,  $Q=0.20$ (left panel) and  $Q=0.50$ (right panel).}
	\label{WECfig}
\end{figure}

Next, we briefly investigate the issue of energy conditions for the stress-energy tensor associated with the charged rotating Hayward black hole (\ref{rotbhtr}). In the frame of a locally nonrotating observer, we consider the following orthonormal tetrad  \cite{Bardeen:1972fi}:
\begin{equation}
e^{(a)}_{\mu}=
\begin{pmatrix}
\sqrt{-(g_{tt}-(g_{t\phi})^2/g_{\phi\phi})} & 0& 0 &0\\
0 & \sqrt{ g_{rr}}& 0 &0\\
0 & 0& \sqrt{ g_{\theta\theta}} &0\\
g_{t\phi}/\sqrt{g_{\phi\phi}} & 0& 0 &\sqrt{ g_{\phi\phi}}\\
\end{pmatrix}.\label{tetrad}
\end{equation}
In this frame the stress-energy tensor is naturally diagonal ($\rho, P_1, P_2, P_3$), whose components read as $T^{(a)(b)}=e^{(a)}_{\mu}e^{(b)}_{\nu}G^{\mu\nu}$ \cite{Balart:2014cga,Toshmatov:2017zpr}. In particular, at the equatorial plane ($\theta=\pi/2$) they have the following form: 
\begin{eqnarray}
T^{(0)(0)}&=& \frac{2r^2m'(r)}{(r^2+a^2)^2}=-T^{(1)(1)},\nonumber\\
T^{(2)(2)}&=&-\frac{r(r^2+a^2)m''(r)+2a^2 m'(r)}{(r^2+a^2)^2}=T^{(3)(3)}.\nonumber
\end{eqnarray}
It is well known that static regular black holes satisfy the weak energy condition ($\rho\geq 0$ and $\rho+P_i\geq 0$) \cite{Bambi:2013ufa,Zaslavskii:2010qz}. In Fig. \ref{WECfig}, we have shown the qualitative behavior of  $\rho+P_2$ as a function of $r$ and $\ell$ for different values of $Q$, nevertheless $\rho>0$ for all possible values. This is evident that the weak energy condition may be violated in the vicinity of the central region ($r<r_c$) of the black hole. This is generic for  all rotating regular black holes (see e.g. \cite{Bambi:2013ufa, Neves:2014aba}). However, despite some energy condition violations, such solutions are important as astrophysical black holes are rotating. However, the violation of the energy condition is very weak (cf. Fig.~\ref{WECfig}), and the region of violation is always shielded by the Cauchy horizon, i.e. $r_c<r_-$ (cf. Fig.~\ref{WECfig} and Table \ref{Table1}). This violation of classical energy conditions is a natural consequence of the fact that the singularity-free metric might incorporate some quantum gravity effects.

\begin{table}[h!]
	\centering
	\begin{tabular}{ |p{1.9cm}|p{1.9cm}|p{1.9cm}|p{1.9cm}|p{1.9cm}|p{1.9cm}| }
	\hline
	
	$Q$&	$r_-$ & $r_+$ & $r^-_{S}$ & $r^+_{S}$ &$ r_c$\\
		\hline
0.00 &0.890837 & 1.64684 & 0.596968 & 1.85464& 0.582197\\
0.10 & 0.90076 & 1.6363 & 0.604679 & 1.84774 & 0.584203\\
0.20 & 0.932269& 1.60285& 0.627961& 1.8266& 0.590225\\
0.30& 0.992836& 1.53874& 0.667561&1.78968& 0.60274\\
0.40& 1.11596& 1.41011&0.725808&1.73368& 0.614348\\
0.434394& 1.26183&1.261283&0.751075&1.70892&0.620121\\
\hline

\end{tabular}
	\caption{Table summarizing the values of $r_-, r_+,r_{S}^-, r_{S}^+$, and $r_c$ for $a=0.50$ and $\ell=0.50$.}\label{Table1}
\end{table}

\section{Conserved quantities}\label{sec4}

The conserved quantities associated with the Killing vectors $\xi_{(t)}^{\mu}$ and  $\xi_{(\phi)}^{\mu}$ can be identified, respectively, as the total mass (effective mass) and angular momentum (effective angular momentum). One can determine these conserved quantities through Komar integrals \cite{Komar:1958wp}, such that they are defined with respect to an observer in the asymptotically flat spacetime unaffected by the spacetime curvature. The metric (\ref{rotbhtr}) is asymptotically Minkowski in a large-$r$ limit, so permits to consider such observers. We consider a spacelike 3-hypersurface $\Sigma_t$ embedded in the 4-dimensional spacetime $\mathcal{M}$, such that the vector $\partial_t$ is orthogonal to the hypersurface. The closed 2-boundary $\partial\mathcal{M}$ of the hypersurface is a constant-$t$ and constant-$r$ surface and spanned only by the $\theta$ and $\phi$ coordinates. The conserved quantities associated with the asymptotically flat spacetime are related to its gravitational Hamiltonian, provided that $\Sigma_t$ asymptotically coincides with the surface of constant-$t$ at the spatial infinity \cite{Chandrasekhar:1992,Wald}. Following a coordinate independent definition of the Komar integral \cite{Komar:1958wp, Cohen:1984ue}, the conserved mass (effective mass of a given spacetime) reads
\begin{equation}
M_{eff}=-\frac{1}{8\pi}\int_{\partial\mathcal{M}} *d\sigma,
\end{equation}
where $\sigma$ corresponds to the timelike one-form            
$$\sigma=\xi_{(t)}{_\mu}dx^{\mu}=g_{t\mu}dx^{\mu}=g_{tt}dt+g_{t\phi}d\phi,$$
and $*d\sigma$ is the dual of two-form $d\sigma$. Since the metric components in Eq.~(\ref{rotbhtr}) are functions of $r$ and $\theta$, the two-form or exterior derivative $d\sigma$ of one-form $\sigma$ reads
\begin{equation}
d\sigma=\frac{\partial g_{tt}}{\partial r}dr\wedge dt + \frac{\partial g_{tt}}{\partial\theta}d\theta\wedge dt + \frac{g_{t\phi}}{\partial r}dr\wedge d\phi + \frac{\partial g_{t\phi}}{\partial \theta}d\theta\wedge d\phi.\label{dual}
\end{equation}
Using Eq.~(\ref{tetrad}), we redefine the tetrad in one-form as
\begin{eqnarray}
e_{(t)}=\sqrt{\frac{\Sigma\Delta}{A}}dt,\;\;\; e_{(r)}=\sqrt{\frac{\Sigma}{\Delta}}dr,\;\;\;
e_{(\theta)}=\sqrt{\Sigma}d\theta,\;\;\; e_{(\phi)}=\frac{2m(r)ar \sin\theta}{\sqrt{\Sigma A}}dt+\sqrt{\frac{A}{\Sigma}}\sin{\theta} d\phi.  
\end{eqnarray}
Then Eq.~(\ref{dual}), in terms of the tetrad, takes the form
\begin{equation}
d\sigma=f e_{(r)}\wedge e_{(t)}  + g e_{(\theta)}\wedge e_{(t)} + h e_{(r)}\wedge e_{(\phi)}  + k e_{(\theta)}\wedge e_{(\phi)}, 
\end{equation}
where 
\begin{eqnarray}
f &=& \left(\frac{A}{\Sigma^2}\right)^{1/2}\frac{\partial g_{tt}}{\partial r} + \frac{2 m(r) a r}{(A\Sigma^2)^{1/2}}\frac{\partial g_{t\phi}}{\partial r},\nonumber\\
g &=& \left(\frac{A}{\Sigma^2\Delta} \right)^{1/2}\frac{\partial g_{tt}}{\partial \theta} + \frac{2 m(r) a r}{(A\Sigma^2\Delta)^{1/2}} \frac{\partial g_{t\phi}}{\partial \theta},\nonumber\\
h &=& \left(\frac{\Delta}{A}\right)^{1/2}\frac{1}{\sin\theta}\frac{\partial g_{t\phi}}{\partial r},\nonumber\\
k &=& \left(\frac{1}{A}\right)^{1/2}\frac{1}{\sin\theta}\frac{\partial g_{t\phi}}{\partial \theta}.
\end{eqnarray}
The dual $*d\sigma$ is a map from two-form to (4-2)-form \cite{Wald}, accordingly it reads
\begin{equation}
*d\sigma=k e_{(t)}\wedge e_{(r)}  - h e_{(t)}\wedge e_{(\theta)} -g e_{(r)}\wedge e_{(\phi)}  + f e_{(\theta)}\wedge e_{(\phi)}.\label{12}
\end{equation}
Here, the integration is performed over the boundary $\partial M$ which is characterized by constant-$t$ and constant-$r$ surface. Using Eq. (\ref{12}) and rewriting it in terms of coordinates, the effective mass reads 
\begin{eqnarray}
M_{eff} &=&\frac{1}{8\pi}\int_{\partial\mathcal{M}}fA^{1/2} \sin\theta d\theta\;d\phi\nonumber\\
&=& m(r)-\frac{(r^2+a^2)}{a}\tan^{-1}\left(\frac{a}{r}\right)m(r)'.\label{Komarmass}
\end{eqnarray}
Inserting the value of mass function $m(r)$ from Eq. (\ref{CHmetric}), we find   
\begin{eqnarray}
M_{eff}= \frac{\left(2Mr-Q^2\right)r^3}{2(r^4+\left(2Mr+Q^2\right)\ell^2)}-\frac{r^2(r^2+a^2)(12M^2r^2l^2-Q^2(3Q^2l^2+4l^2Mr+r^4))}{2a(r^4+\left(2Mr+Q^2\right)\ell^2)^2}\tan^{-1}\left(\frac{a}{r}\right)\label{KMass}
\end{eqnarray}

At the asymptotic spatial infinity, the observed mass yields $\lim_{r\rightarrow\infty}M_{eff}=M$, whereas in the limit $\ell=0$, it reads
\begin{equation}
M_{eff}=M-\frac{Q^2}{2r}-\frac{Q^2(r^2+a^2)}{2ar^2}\tan^{-1}\left(\frac{a}{r}\right),
\end{equation} 
 which is the $M_{eff}$ for the Kerr-Newman black hole \cite{Modak:2010fn}. Furthermore, in the limit $a=0$, we obtain the effective mass of the static charged Hayward black hole
\begin{equation}
M_{eff}=\frac{(2Mr-Q^2)r^3-(8Mr-3Q^2)r^4}{2 (r^4+(2Mr+Q^2)l^2)}  + \frac{(2Mr-Q^2)(4r^3+2Ml^2)r^5}{2 (r^4+(2Mr+Q^2)l^2)^2}.
\end{equation} 
Now, we evaluate the effective angular momentum by using the Komar integral for the spcaelike Killing vector $\xi_{(\phi)}^{\mu}$, which reads \cite{Komar:1958wp}
 \begin{equation}
 J_{eff}=\frac{1}{16\pi}\int_{\partial\mathcal{M}} *d\eta.\label{angmom2}
 \end{equation}
Here, the prefactor $1/{16\pi}$ is chosen properly to obtain the correct value at the asymptotic infinity, and $d\eta$ is a two-form for the spacelike one-form $\eta$ defined as
$$\eta=\xi_{(\phi)}{_\mu}dx^{\mu}=g_{\phi\mu}dx^{\mu}=g_{t\phi}dt+g_{\phi\phi}d\phi,$$
\begin{equation}
d\eta=\frac{\partial g_{t\phi}}{\partial r}dr\wedge dt + \frac{\partial g_{t\phi}}{\partial\theta}d\theta\wedge dt + \frac{\partial g_{\phi\phi}}{\partial r}dr\wedge d\phi + \frac{\partial g_{\phi\phi}}{\partial \theta}d\theta\wedge d\phi.\label{dual1}
\end{equation}
Hence, we obtain the dual to two-form $d\eta$
\begin{equation}
*d\eta=k_1 e_{(t)}\wedge e_{(r)}  - h_1 e_{(t)}\wedge e_{(\theta)} -g_1 e_{(r)}\wedge e_{(\phi)}  + f_1 e_{(\theta)}\wedge e_{(\phi)},\label{angmom1}
\end{equation}
where
\begin{eqnarray}
f_1 &=& \left(\frac{A}{\Sigma}\right)^{1/2}\frac{\partial g_{t\phi}}{\partial r} + \frac{2 m(r) a r}{(A\Sigma^2)^{1/2}}\frac{\partial g_{\phi\phi}}{\partial r}\nonumber,\\
g_1 &=& \left(\frac{A}{\Sigma^2\Delta} \right)^{1/2}\frac{\partial g_{t\phi}}{\partial \theta} + \frac{2 m(r) a r}{(A\Sigma\Delta)^{1/2}} \frac{\partial g_{\phi\phi}}{\partial \theta}\nonumber,\\
h_1 &=& \left(\frac{\Delta}{A}\right)^{1/2}\frac{1}{\sin\theta}\frac{\partial g_{\phi\phi}}{\partial r}\nonumber,\\
k_1 &=& \left(\frac{1}{A}\right)^{1/2}\frac{1}{\sin\theta}\frac{\partial g_{\phi\phi}}{\partial \theta}.
\end{eqnarray}

\begin{figure}
	 \begin{tabular}{c c}
\includegraphics[scale=0.66]{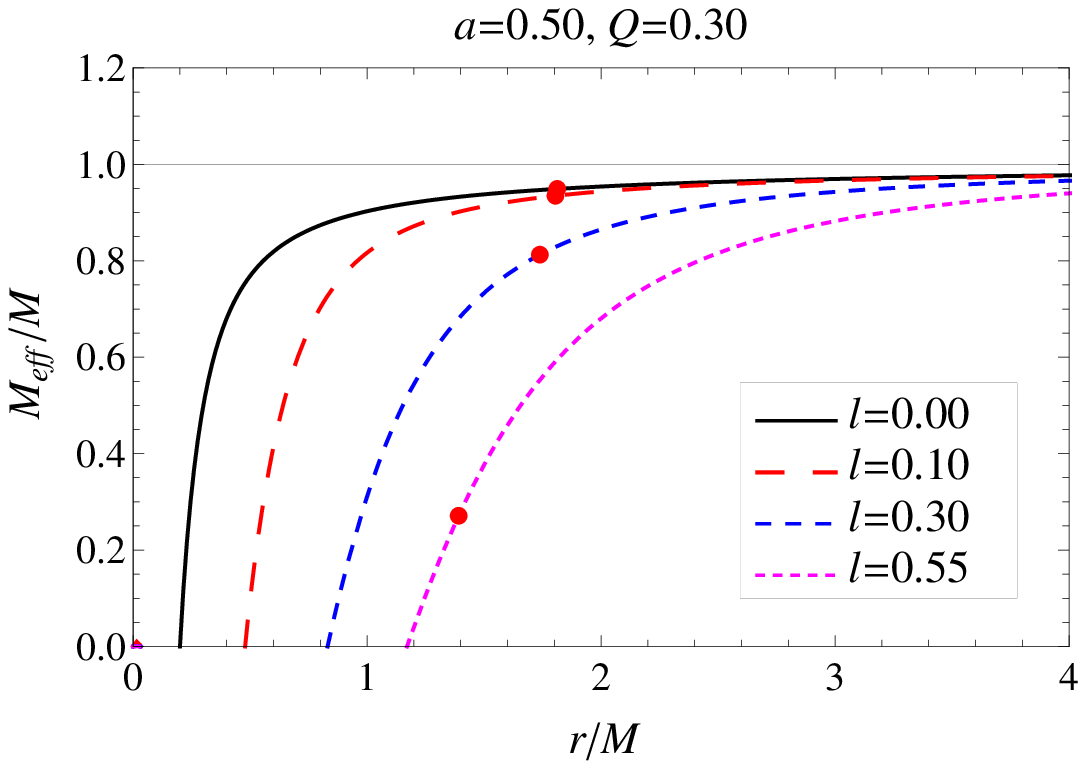}&
\includegraphics[scale=0.66]{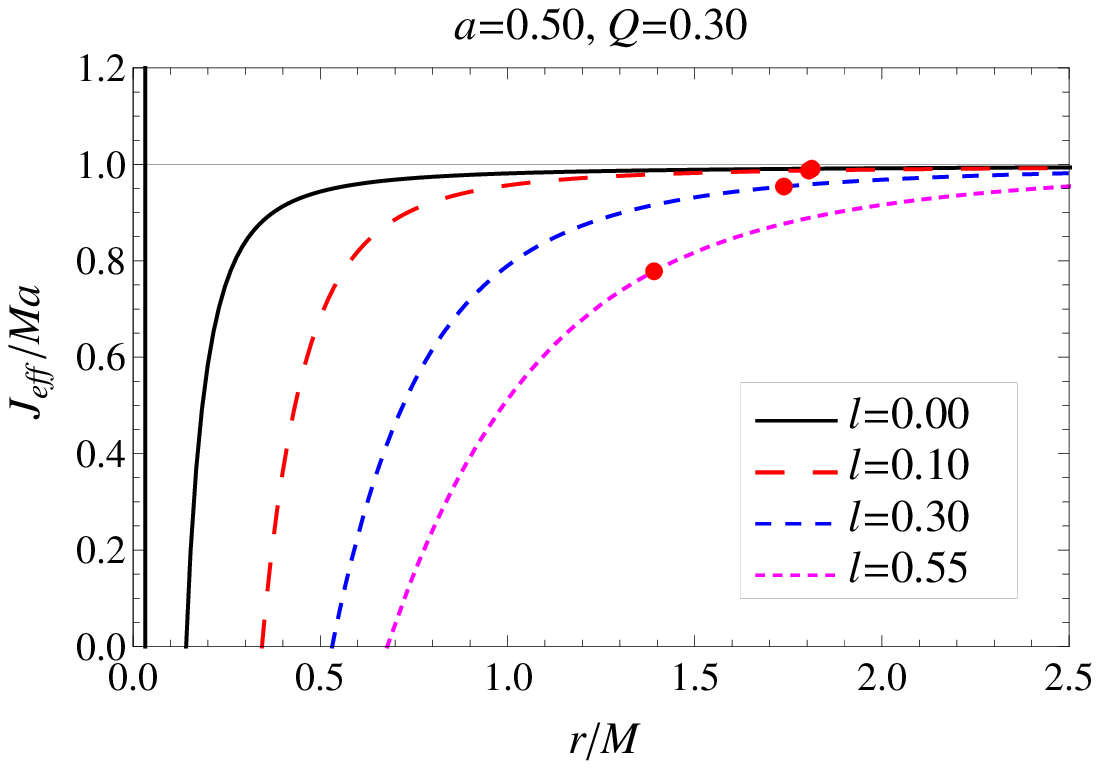}
     \end{tabular}
 	\caption{The behavior of the effective mass and angular momentum vs $r$ for different values of the parameters. The black solid curve corresponds to the Kerr-Newman black hole and red dots in each curve denote  the locations of the event horizon. }
 \label{Komar}
\end{figure}

Using Eq.~(\ref{angmom1}), we find that Eq.~(\ref{angmom2}) reduces to 
\begin{eqnarray}
J_{eff} &=&-\frac{1}{16\pi}\int_{\partial\mathcal{M}}f_1A^{1/2} \sin\theta d\theta\;d\phi\nonumber\\
&=& m(r)a+\frac{(r^2+a^2)m(r)'r}{2a}-\frac{1}{2}\frac{(r^2+a^2)^2}{a^2}m(r)'\tan^{-1}\left(\frac{a}{r}\right).\label{Komarang}
\end{eqnarray}
Equation (\ref{Komarang}) with the charged Hayward mass function (\ref{CHmetric}) gives the Komar angular momentum in terms of $M, a, Q, l$, which in the limit $\ell=0$ yields 
\begin{equation}
J_{eff}=Ma+\frac{(r^2-a^2)arQ^2}{4a^2r^2}-\frac{Q^2(r^2+a^2)^2}{4a^2r^2}\tan^{-1}\left(\frac{a}{r}\right),
\end{equation}
which can be identified as the effective angular momentum for the Kerr-Newman spacetime. In the large $r$ limit the effective angular momentum $J_{eff}$ (\ref{Komarang}) reduces to the value $Ma$. 
The expressions in Eqs.(\ref{Komarmass}) and (\ref{Komarang}) account for the effective mass and angular momentum measured within the 2-sphere of radius $r$. The normalized effective mass and angular momentum against $r$ for various values of $\ell$ are depicted in Fig. \ref{Komar}, from which it can be clearly inferred that the values of effective quantities decrease gradually with decreasing $r$. It is worth mentioning that the presence of charge $Q$ reduces the values of the effective mass and angular momentum, i.e., $M_{eff}/M\leq 1$ and $J_{eff}/Ma\leq 1$. Moreover, at a fixed radial coordinate the normalized values of effective quantities for the regular black hole ($\ell\neq 0$) are less than that for a singular black hole ($\ell=0$). Indeed, the effect of the nonzero value of $\ell$ is appreciable only near the event horizon $r_+$, such that at long distances from $r_+$ the effect of $\ell$  is washed away, i.e., $M_{eff}/M=1$ and $J_{eff}/Ma=1$ at large$-r$ (cf. Fig. \ref{Komar}). Therefore, $M_{eff}$ and $J_{eff}$ are always smaller than their asymptotic values. 
 Apart from the Killing vectors $\xi_{(t)}^{\mu}$ and  $\xi_{(\phi)}^{^\mu}$, their linear combinations also produce the isometries of metric. In particular, $\chi^{\mu}=\xi_{(t)}^{\mu}+\Omega \xi_{(\phi)}^{\mu}$, which is a global timelike vector (for $r>r_+$) is a generator of the Killing horizon as well, where $\Omega$ turns out to be the angular velocity at the event horizon. Therefore, we can identify a conserved quantity associated with $\chi_{\mu}$. Following the definitions of Komar integrals, we find
 \begin{eqnarray}
K &=& -\frac{1}{8\pi}\int_{\partial\mathcal{M}} *d\chi,\nonumber\\
&=& M_{eff}-2 \Omega J_{eff},\label{KomarKV}
\end{eqnarray}
with 
\begin{equation}
\Omega=\frac{a}{r_+^2+a^2}.
\end{equation}
Substituting the expression for $M_{eff}$ and $J_{eff}$ from Eqs. (\ref{Komarmass}) and (\ref{Komarang}) into Eq.(\ref{KomarKV}), we find    the Komar conserved quantity at the event horizon 
\begin{equation}
K=(1-\frac{2a^2}{r_+^2+a^2})m(r)-m(r)'r,
\end{equation}
which in the limit $l\to 0$ retains the following value 
\begin{equation}
K=\frac{M(r_+^2-a^2)}{r_+^2+a^2}-\frac{r_+Q^2}{r_+^2+a^2}.
\end{equation}
This  is consistent with the value for the Kerr-Newman black hole \cite{Modak:2010fn}. 

\section{Black hole shadow}\label{sec5}

The EHT was set up for imaging  shadows of supermassive black holes like M87* and Sgr A* \cite{EHT}. Recently, the EHT has released the first image of M87*, which is in accordance with the shadow of a Kerr black hole as predicted by general relativity \cite{Akiyama:2019cqa,Akiyama:2019eap}. It turns out that the Reissner-Nordstr\"{o}m black hole with a significant charge also agrees with the observational data, and provides even a better fit when compared with the Schwarzschild black hole \cite{Zakharov:2014lqa}. The shadow of a black hole is the boundary of photon capture orbits and scattering orbits \cite{Chandrasekhar:1992}, and hence in the following let us consider photon orbits in the background of the charged rotating Hayward black holes. Isometries along $\partial_t$ and $\partial_{\phi}$ allow one to define the conserved energy $\mathcal{E}$ and angular momentum $\mathcal{L}$ along the geodesics
\begin{eqnarray}
-\mathcal{E}=g_{\mu\nu}\xi^{\mu}_{(t)}u^{\nu}=g_{tt}\dot{t}+g_{t\phi}\dot{\phi},\nonumber\\
\mathcal{L}=g_{\mu\nu}\xi^{\mu}_{(\phi)}u^{\nu}=g_{t\phi}\dot{t}+g_{\phi\phi}\dot{\phi}.
\end{eqnarray} 
Solving these equations leads to the following geodesic equations in the first-order differential form
\begin{eqnarray}
\Sigma \frac{dt}{d\tau}&=&\frac{r^2+a^2}{\Delta}\left({\cal E}(r^2+a^2)-a{\cal L}\right)  -a(a{\cal E}\sin^2\theta-{\mathcal {L}})\ ,\label{tuch}\\
\Sigma \frac{d\phi}{d\tau}&=&\frac{a}{\Delta}\left({\cal E}(r^2+a^2)-a{\cal L}\right)-\left(a{\cal E}-\frac{{\cal L}}{\sin^2\theta}\right)\ .\label{phiuch}
\end{eqnarray}
Following the Hamilton-Jacobi equation for geodesic motions \cite{Carter:1968rr}
\begin{eqnarray}
\label{HmaJam}
\frac{\partial S}{\partial \tau} = -\frac{1}{2}g^{\alpha\beta}\frac{\partial S}{\partial x^\alpha}\frac{\partial S}{\partial x^\beta},
\end{eqnarray}
and Carter's  separability prescription \cite{Carter:1968rr}, we choose the action as
\begin{eqnarray}
S=\frac12 {m_0}^2 \tau -{\cal E} t +{\cal L} \phi +S_r(r)+S_\theta(\theta) \label{action},
\end{eqnarray}
where $\tau$ is the affine parameter along the geodesics. Then, we obtain the geodesic equation of motion for $r$ and $\theta$ coordinates
\begin{eqnarray}
\Sigma \frac{dr}{d\tau}&=&\pm\sqrt{\mathcal{R}(r)}\ ,\label{r}\\
\Sigma \frac{d\theta}{d\tau}&=&\pm\sqrt{\Theta(\theta)}\ ,\label{th}
\end{eqnarray}
with 
\begin{eqnarray}\label{06}
\mathcal{R}(r)&=&\left((r^2+a^2){\cal E}-a{\cal L}\right)^2-\Delta ((a{\cal E}-{\cal L})^2+{\cal K}),\quad \\ 
\Theta(\theta)&=&{\cal K}-\left(\frac{{\cal L}^2}{\sin^2\theta}-a^2 {\cal E}^2\right)\cos^2\theta,\label{theta0}
\end{eqnarray}
where ${\cal K}$ stands for Carter's constant of motion \cite{Carter:1968rr}. Equations (\ref{tuch}), (\ref{phiuch}) along with Eqs.(\ref{r}) and (\ref{th}) govern the null geodesics around the black hole. In principle, depending on the dimensionless impact parameters $\eta\equiv{\cal K}/\mathcal{E}^2$ and $\xi\equiv\mathcal{L}/\mathcal{E}$ photons may undergo three different kinds of geodesics,  namely, scattering, spherical orbits, and plunging orbits \cite{Chandrasekhar:1992}. Those photons which cross the event horizon and eventually fall into the black hole account for the dark region of the shadow against the bright background, whereas scattered photons reach the observer. On the other hand, the unstable photon orbits define the shadow boundary, which demarcates the dark and bright regions \cite{Synge:1966,Luminet:1979,Bardeen}. Thus the shadow comprises the optical appearance of the black hole. Photons undergoing the unstable orbits experience turning points in their radial motions, such that they form  a photon region around the black hole filled with spherical photon orbits of constant radii. This demands a local maximum of potential
\begin{eqnarray}
 \left.\mathcal{R}\right|_{(r=r_p)}=\left.\frac{\partial \mathcal{R}}{\partial r}\right|_{(r=r_p)}=0 \quad \text{and} \quad \left.\frac{\partial^2 \mathcal{R}}{\partial r^2}\right|_{(r=r_p)}> 0,
\end{eqnarray}
which gives the locus of the critical impact parameters $\eta_s$ and $\xi_s$, where $r_p$ is the photon orbit radius. This together with the condition $\Theta(\theta)\geq 0$ characterizes the photon region. A black hole shadow can be visualized as a projection from the celestial sphere to the observer's image plane, so we define the celestial coordinate $\alpha$ and $\beta$ as \cite{Hioki:2009na} 
\begin{equation}
\alpha=\lim_{r_*\rightarrow\infty}\left(-r_*^2 \sin{\theta_0}\frac{d\phi}{d{r}}\right),\quad  \beta=\lim_{r_*\rightarrow\infty}r_*^2\frac{d\theta}{dr},\label{celestial}
\end{equation}
where $r_*$ is the distance between the observer and  black hole, and $\theta_0$ is the inclination angle between the line of sight of the observer and the rotational axis of the black hole. We consider that observer is at a far distance from the black hole and lies on the $\theta_0=\pi/2$ plane. Using geodesic Eqs. (\ref{phiuch})-(\ref{r}) and celestial coordinate (\ref{celestial}), we obtain 
\begin{figure}
	\begin{tabular}{c c}
		\includegraphics[scale=0.9]{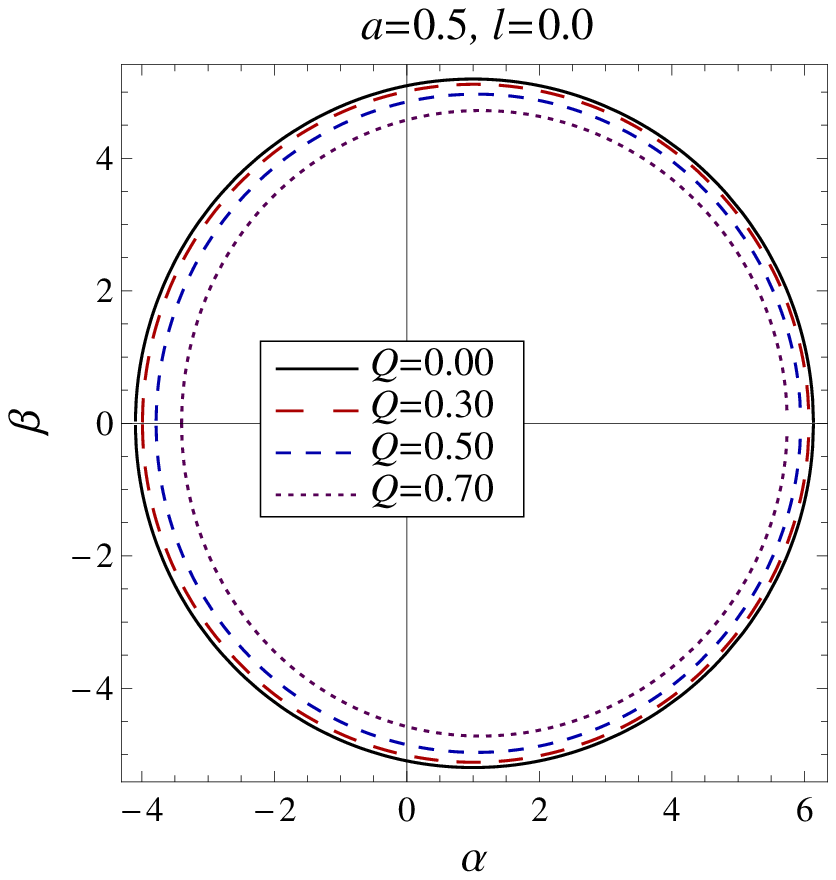}&
		\includegraphics[scale=0.9]{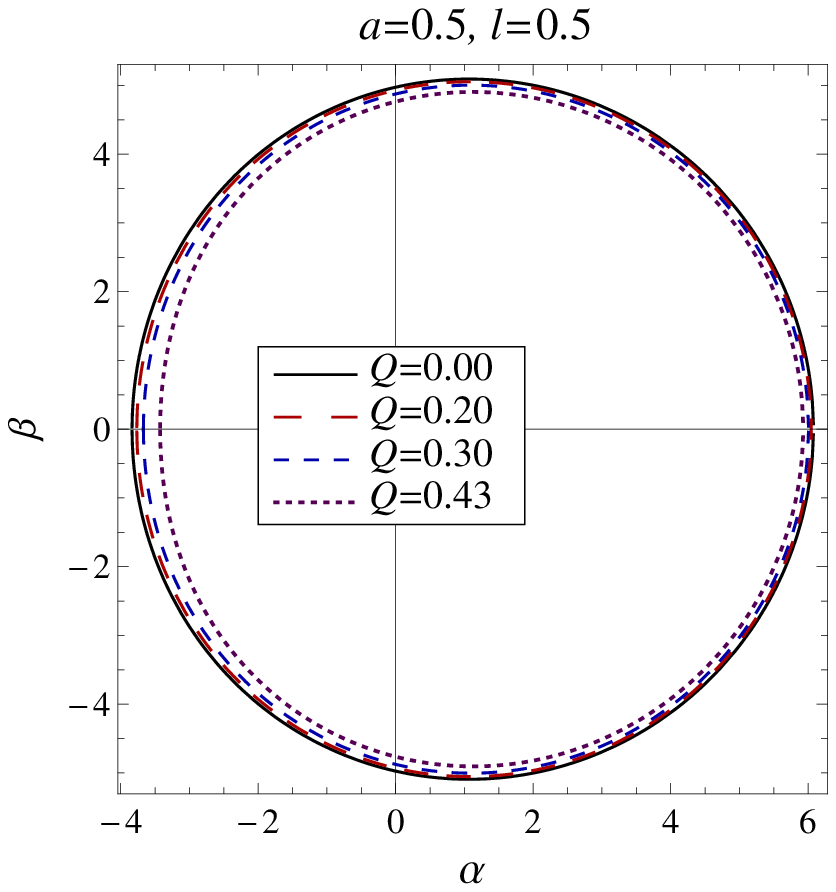}\\
		\includegraphics[scale=0.9]{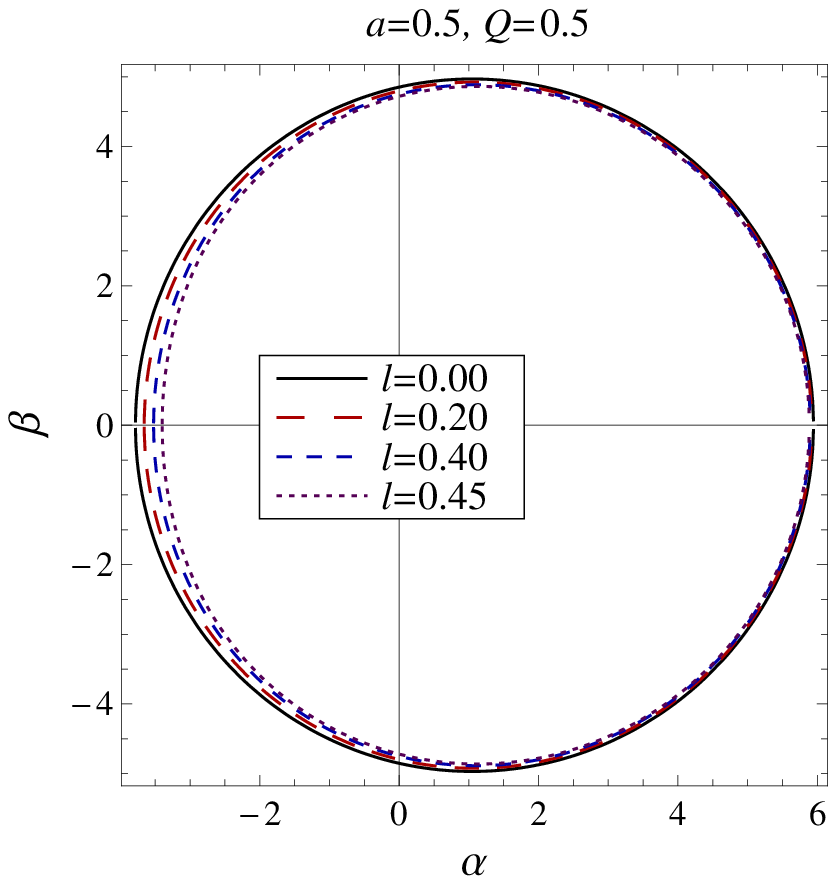}&
		\includegraphics[scale=0.9]{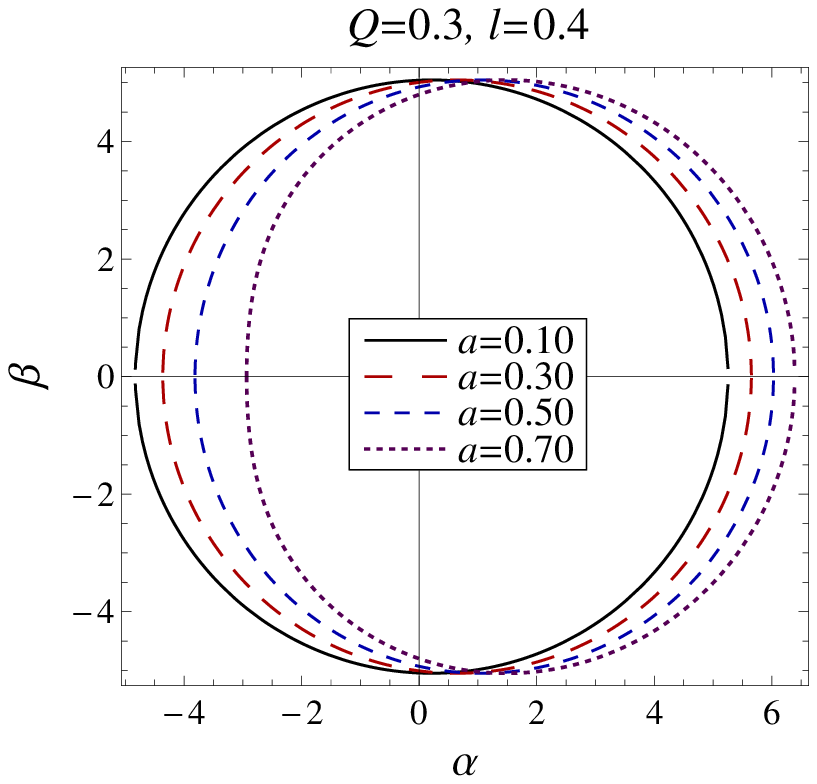}
	\end{tabular}
		\caption{The silhouette of the charged rotating Hayward black hole with varying parameters. }
\label{shadow}
\end{figure}

\begin{eqnarray}
\alpha&=&-\xi_s= \frac{r_p^3 (-r_p^3 + m(r_p) (4 a^2 + 6 r_p^2 - 9 m(r_p) r_p ) - 2 r_p (2 a^2 + r_p^2 - 3 m(r_p) r_p) m(r_p)' - r_p^3 m(r_p)'^2)}{a^2 (m(r_p) + r_p (-1 + m(r_p)'))^2},\nonumber\\
\beta&=&\sqrt{\eta_s}=\frac{(a^2 - 3 r_p^2) m(r_p) + r_p (a^2 + r_p^2) (1 + m(r_p)')}{a (m(r_p)+ r_p (-1 + m(r_p)'))}.\label{alpha}
\end{eqnarray}

\begin{figure}
	\begin{tabular}{c c}
		\includegraphics[scale=0.7]{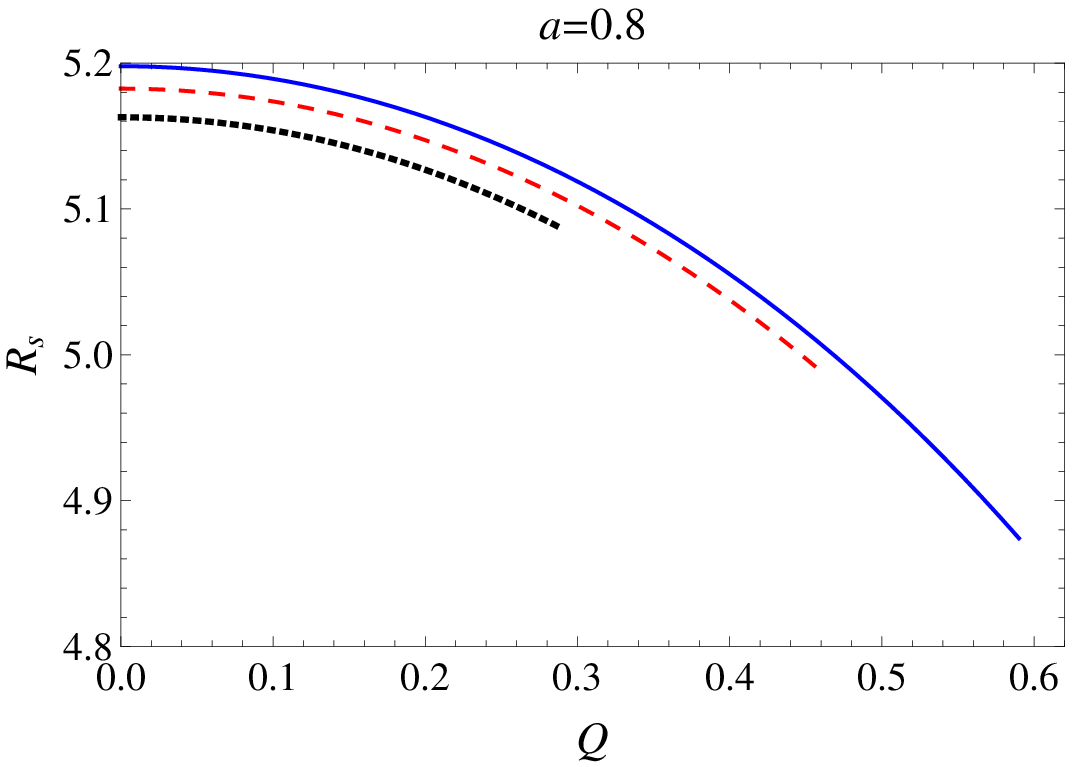}&
		\includegraphics[scale=0.7]{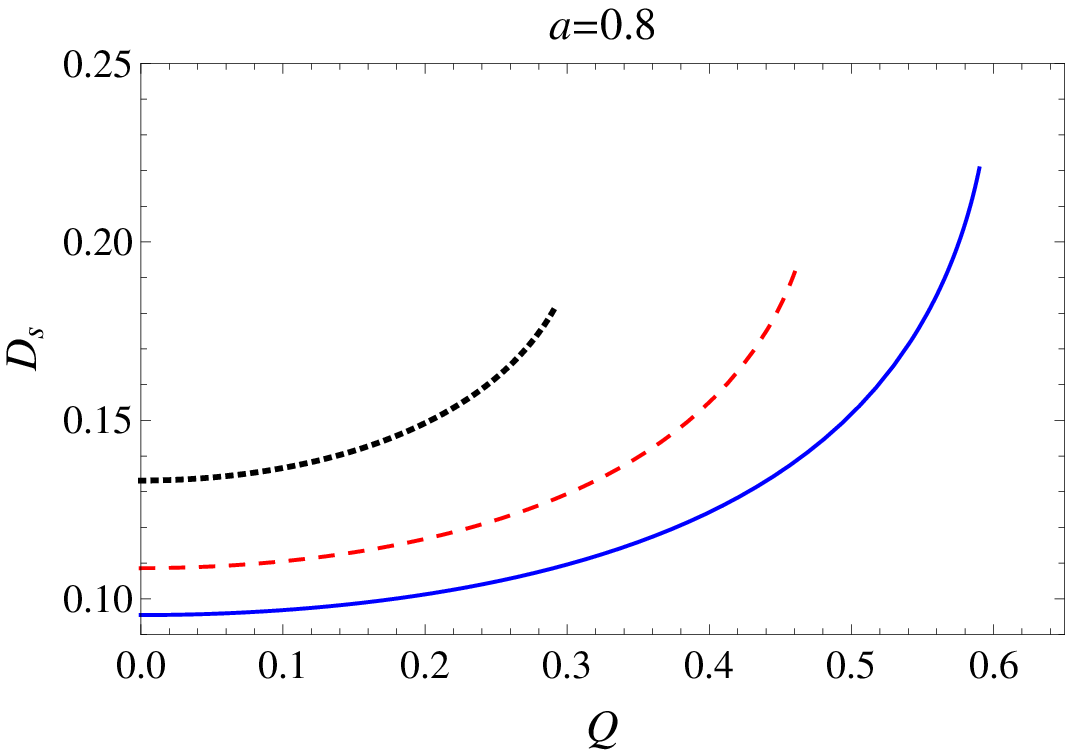}
	\end{tabular}
	\caption{The shadow observables $R_s$ and $\delta_s$ of the charged rotating Hayward black hole with varying $Q$ in geometrized units, (blue solid curve) $\ell=0.0$, (red dashed curve) $\ell=0.2$, and (black dotted curve) $\ell=0.3$. }
	\label{observables}
\end{figure}

The contour of $\alpha$ and $\beta$ in Eq.~(\ref{alpha}) delineate the shadow for the charged rotating Hayward black hole, which is depicted in Fig. \ref{shadow} for varying parameters: the shadow corresponds to the region inside each closed curve. A comparison of the charged rotating Hayward black hole shadow with that for the Kerr-Newman ($\ell=0$) and rotating Hayward ($Q=0$) is also shown. Further, in order to characterize the apparent shadow, we define two observables,  namely, shadow radius ($R_s$) and distortion parameter ($\delta_s$) \cite{Hioki:2009na}. We approximate the shadow periphery by a referenced circle, that coincide at the top, bottom, and extreme right edges of the shadow. Indeed, $R_s$ is the radius of the circle, while $\delta_s$ is the measure of deformation of the shadow from a circle. The prograde photons experience a comparatively different effective potential than the retrograde one; this in turn eventually leads to the apparent distortion in the rotating black hole shadow. These observables are defined as \cite{Hioki:2009na}
\begin{equation}
R_s=\frac{(\alpha_t-\alpha_r)^2+\beta_t^2}{2|\alpha_t-\alpha_r|},\;\;\;\;\   \delta_s=\frac{\tilde{\alpha}_l-\alpha_l}{R_s}.
\end{equation} 
Here, $(\alpha_t,\beta_t), (\alpha_r,\beta_r), (\alpha_l,\beta_l)$ are, respectively, the coordinates of the shadow vertices at top, right and left edges, while $(\tilde{\alpha}_l, \tilde{\beta}_l)$ is the coordinate of the left edge of the referenced circle \cite{Abdujabbarov:2016hnw}.

\begin{figure}
	\begin{tabular}{c c}
	\includegraphics[scale=0.9]{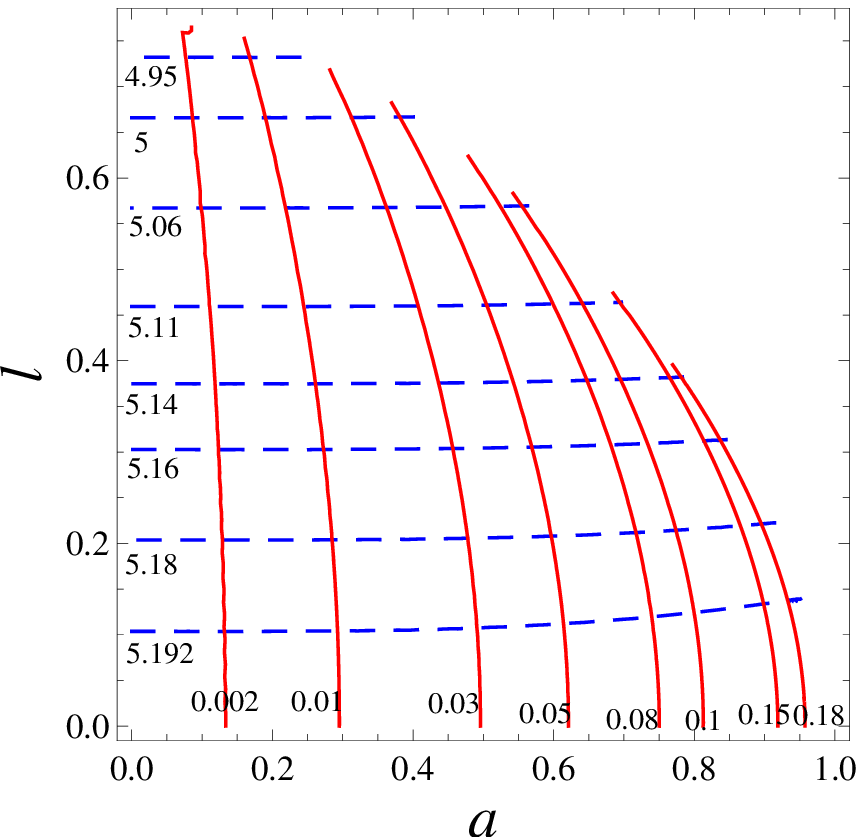} &
	\includegraphics[scale=0.9]{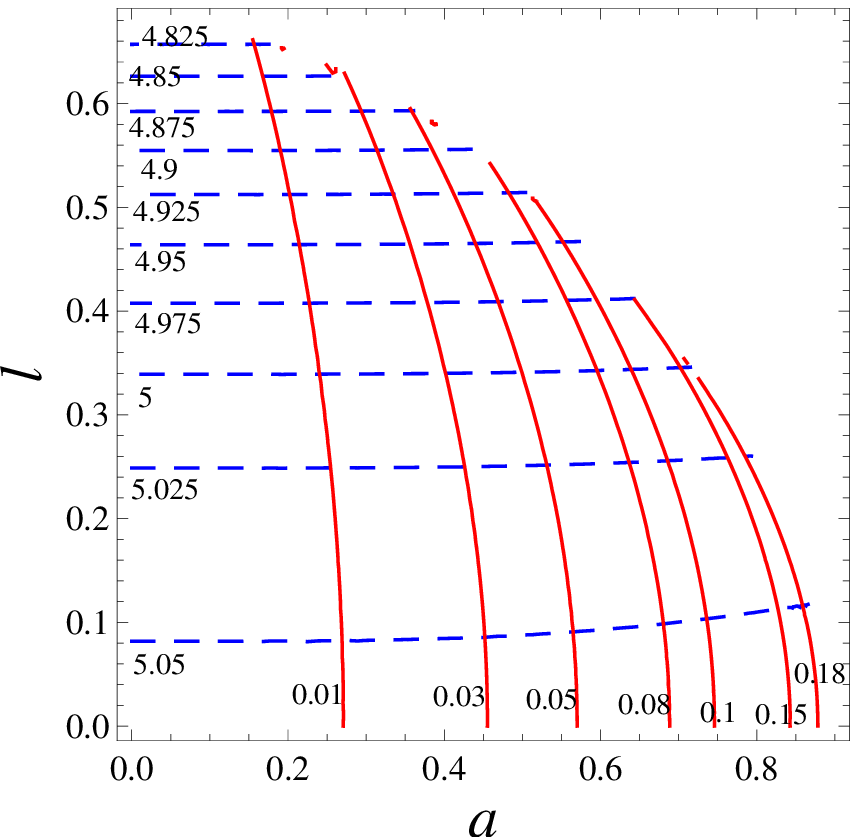}
		\end{tabular}
	\caption{Observables $R_s$ and $\delta_s$ are plotted in the ($a,\ell$) plane for $Q=0.0$ (left panel) and $Q=0.4$ (right panel). Blue dashed curves correspond to  constant $R_s$ and red solid curves are for constant $\delta_s$. The point of intersection determines the values of the black hole parameters.}\label{RSDS}
\end{figure}

\begin{figure}
	\begin{tabular}{c c}
		\includegraphics[scale=0.87]{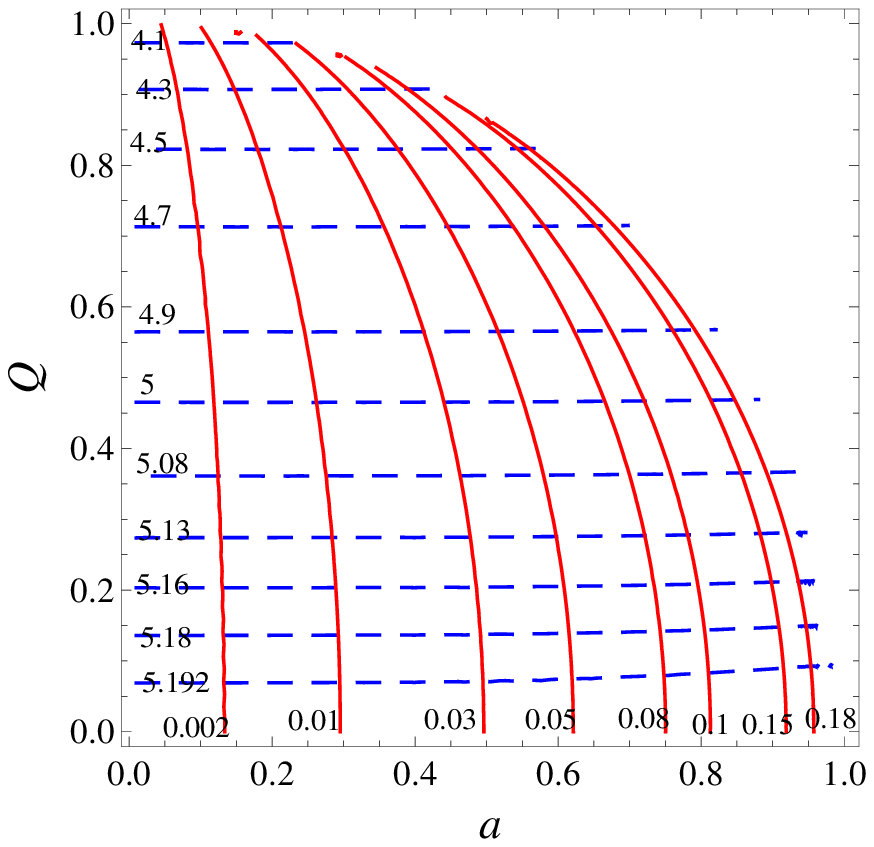} &
		\includegraphics[scale=0.87]{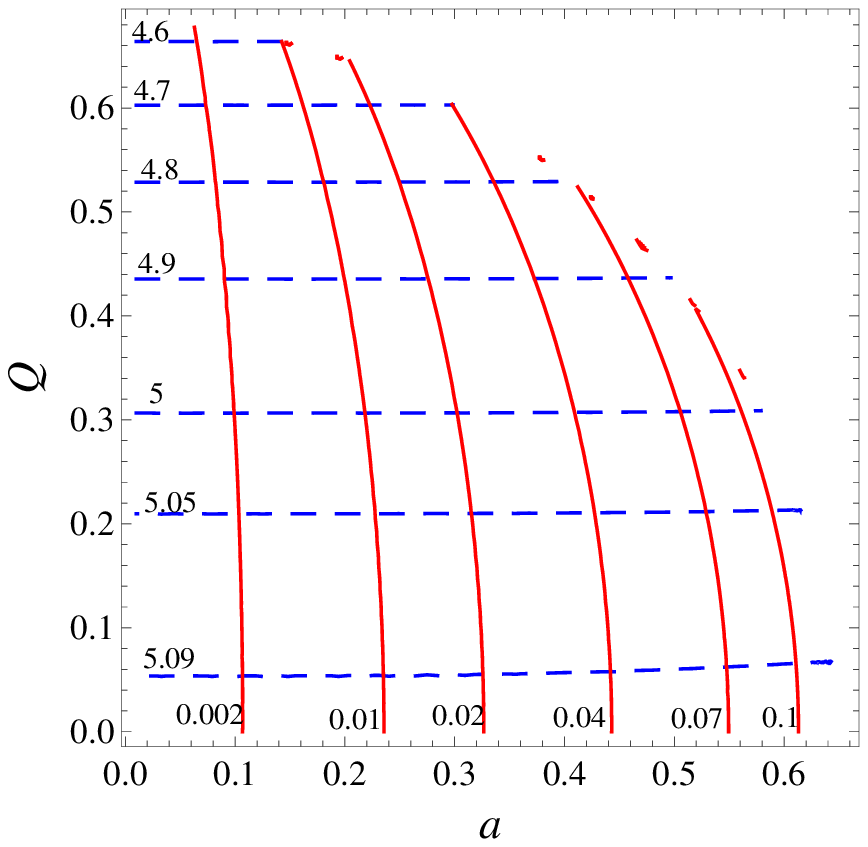}\\
	\end{tabular}
	\caption{Observables $R_s$ and $\delta_s$ are plotted in the ($a,Q$) plane for $\ell=0.0$ (\textit{left panel}) and $\ell=0.5$ (\textit{right panel}). Blue dashed curves correspond to constant $R_s$ and red solid curves are for constant $\delta_s$. }\label{RSDSfig}
\end{figure}

The presence of the charge has a profound influence on the apparent shape and size of the shadow, as increasing $Q$ gradually decreases the shadow size whereas it increases the distortion (cf. Fig. \ref{observables}). Moreover, the distortion in the shadow grows rapidly for the near extremal values of black hole parameters. Similarly, for fixed values of $a$ and $Q$, increasing $\ell$ reduces the size of the shadow and enhances the distortion, viz., the shadow of the charged rotating Hayward black hole is smaller and more distorted than the corresponding Kerr-Newman black hole shadow ($\ell=0$). The blue solid curve in Fig. \ref{observables} corresponds to the Kerr-Newman black hole. We have plotted these shadow observables in the ($a, \ell$) and ($a, Q$) planes, respectively, in Figs. \ref{RSDS} and \ref{RSDSfig}. This is evident from figures that curves of constant $R_s$ and $\delta_s$ intersect at unique points that precisely determine the black hole parameters. 

The parameters of the charged rotating Hayward black hole are expected to be adequately constrained from the recent observation of the black hole shadow by the EHT Collaboration. The observed image of the M87* black hole is consistent with the Kerr black hole shadow as predicted by the general relativity. Nevertheless, Kerr-deviated black holes can also fit with the observational data \cite{Akiyama:2019eap,Bambi:2019tjh,Banerjee:2019nnj}. The observation inferred that the shadow is nearly circular and the deviation from circularity in terms of the root-mean-square (RMS) distance from the shadow average radius is $\Delta C\leq 10\%$. We define the average radius of the shadow as \cite{Johannsen:2010ru}
\begin{eqnarray}
\bar{R}&=&\frac{1}{2\pi}\int_{0}^{2\pi} R(\varphi) d\varphi,\nonumber\\
R(\varphi)&=& \sqrt{(\alpha-\alpha_G)^2+(\beta-\beta_G)^2},
\end{eqnarray}
where $(\alpha_G,\beta_G)$ is the geometric center of the shadow; $\alpha_G$ is the horizontal displacement, and $\beta_G$ is the vertical displacement 
\begin{equation}
\alpha_G=\frac{\mid \alpha_{max}+\alpha_{min}\mid}{2},\qquad \beta_G=\frac{\mid \beta_{max}+\beta_{min}\mid}{2},
\end{equation}
due to the shadow symmetry along the $\alpha$ axes $\beta_G=0$, and $\varphi$ determines the angle along the shadow boundary from the $\alpha$ axes
\begin{equation}
\varphi\equiv \tan^{-1}\left(\frac{\beta}{\alpha-\alpha_G}\right).
\end{equation}
The deviation from the circularity is defined as \cite{Johannsen:2010ru,Johannsen:2015qca}
\begin{equation}
\Delta C=2\sqrt{\frac{1}{2\pi}\int_0^{2\pi}\left(R(\varphi)-\bar{R}\right)^2d\varphi}.
\end{equation}

\begin{figure}
	\begin{tabular}{c c}
\includegraphics[scale=0.7]{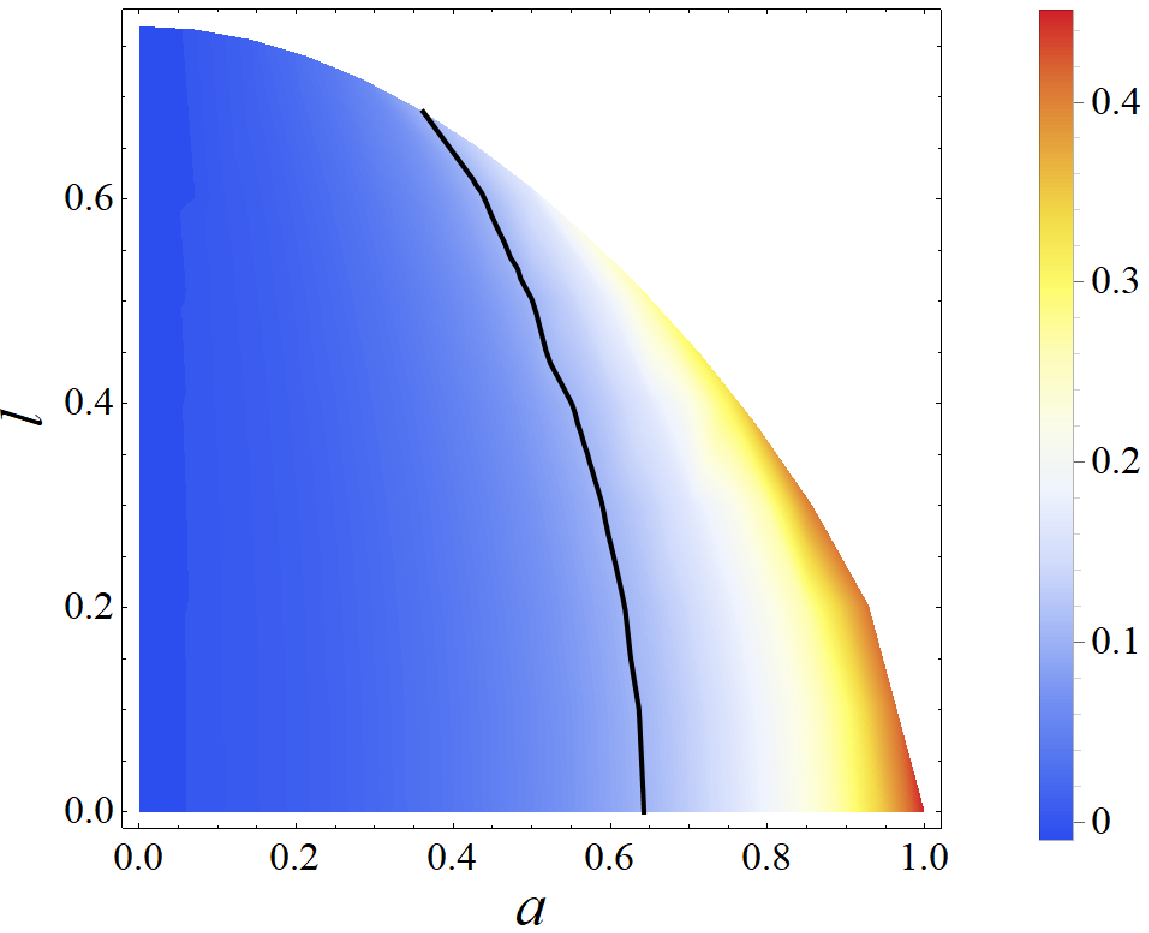} &
\includegraphics[scale=0.7]{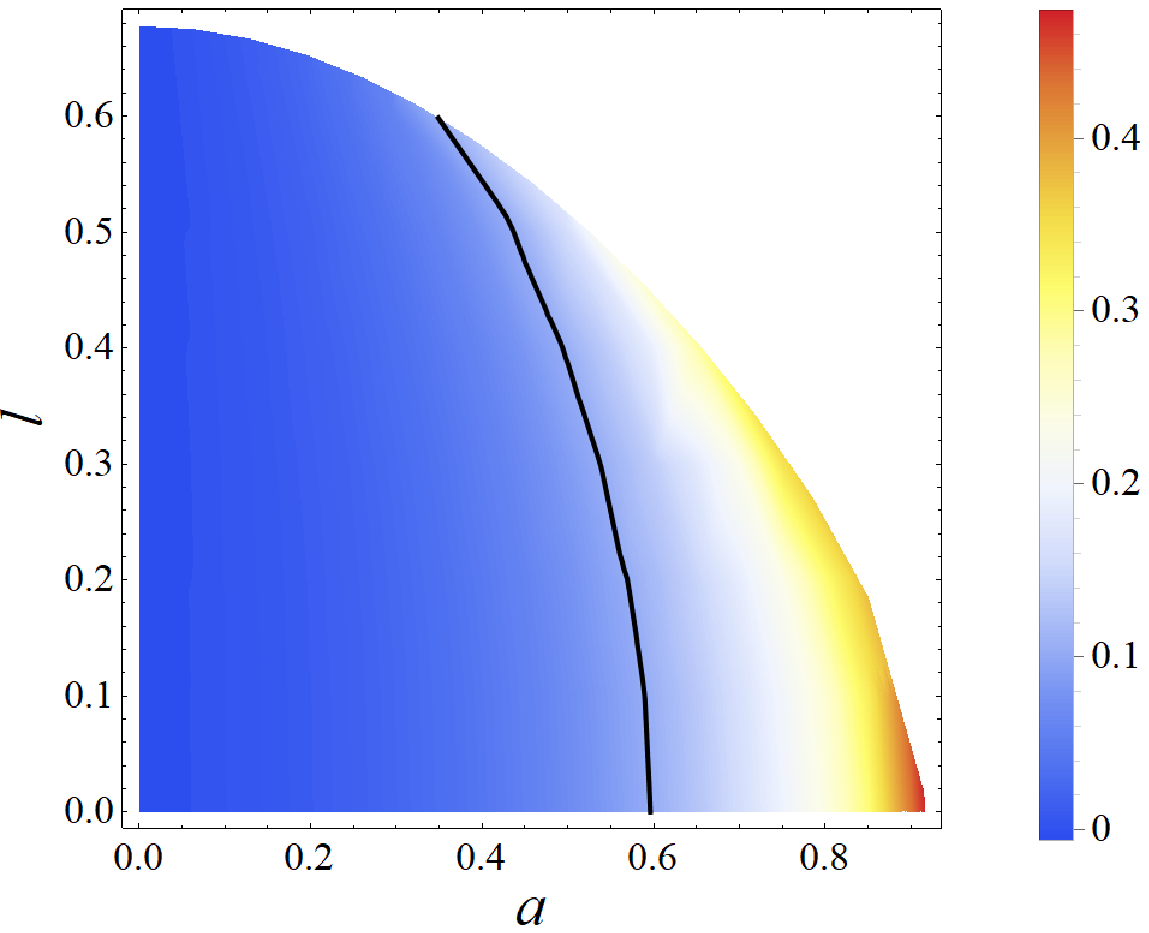}
	\end{tabular}
\caption{The deviation from circularity $\Delta C$ as a function of $a$ and $l$ for $Q=0.0$ (left panel) and $Q=0.40$ (right panel). The black solid line corresponds to $\Delta C=0.10$.  }\label{M87}
\end{figure}
\begin{figure}[h!]
	\begin{tabular}{c c}
		\includegraphics[scale=0.7]{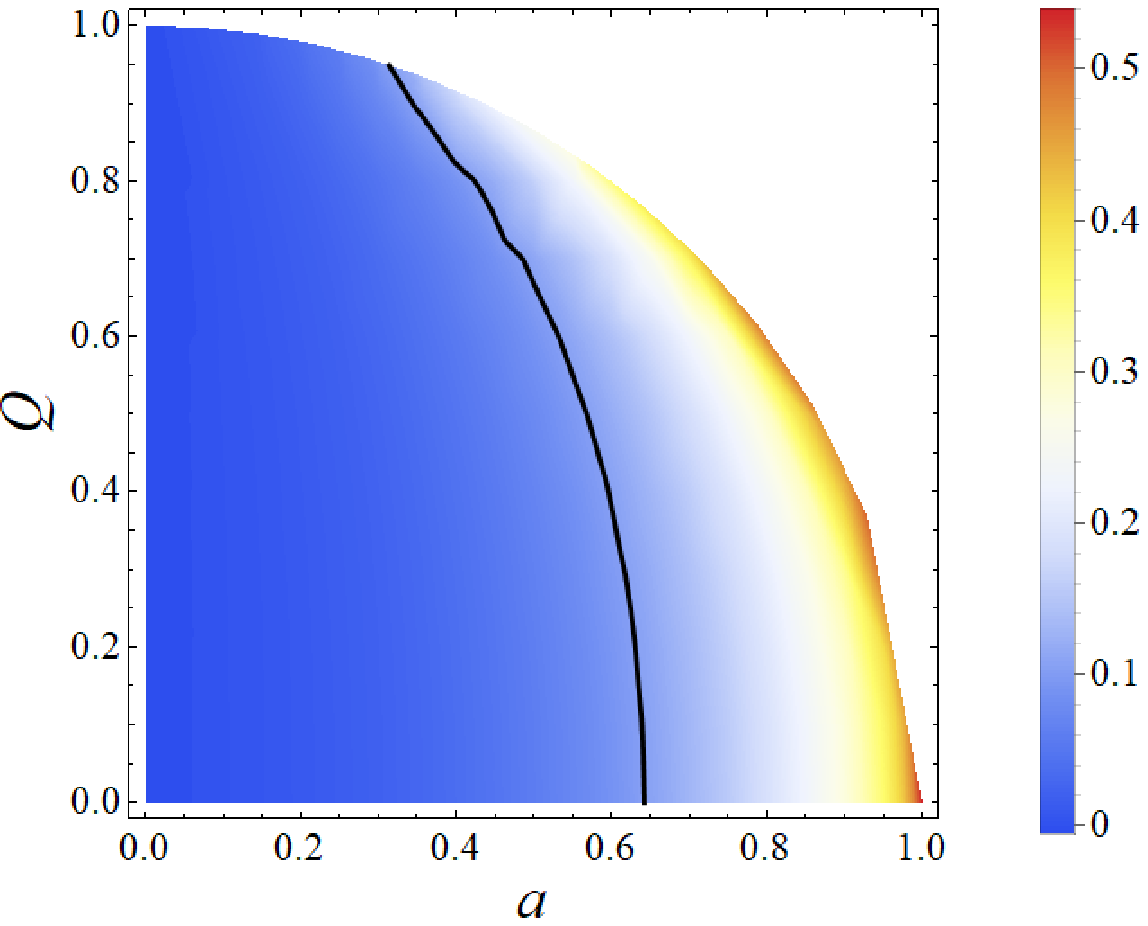} &
		\includegraphics[scale=0.7]{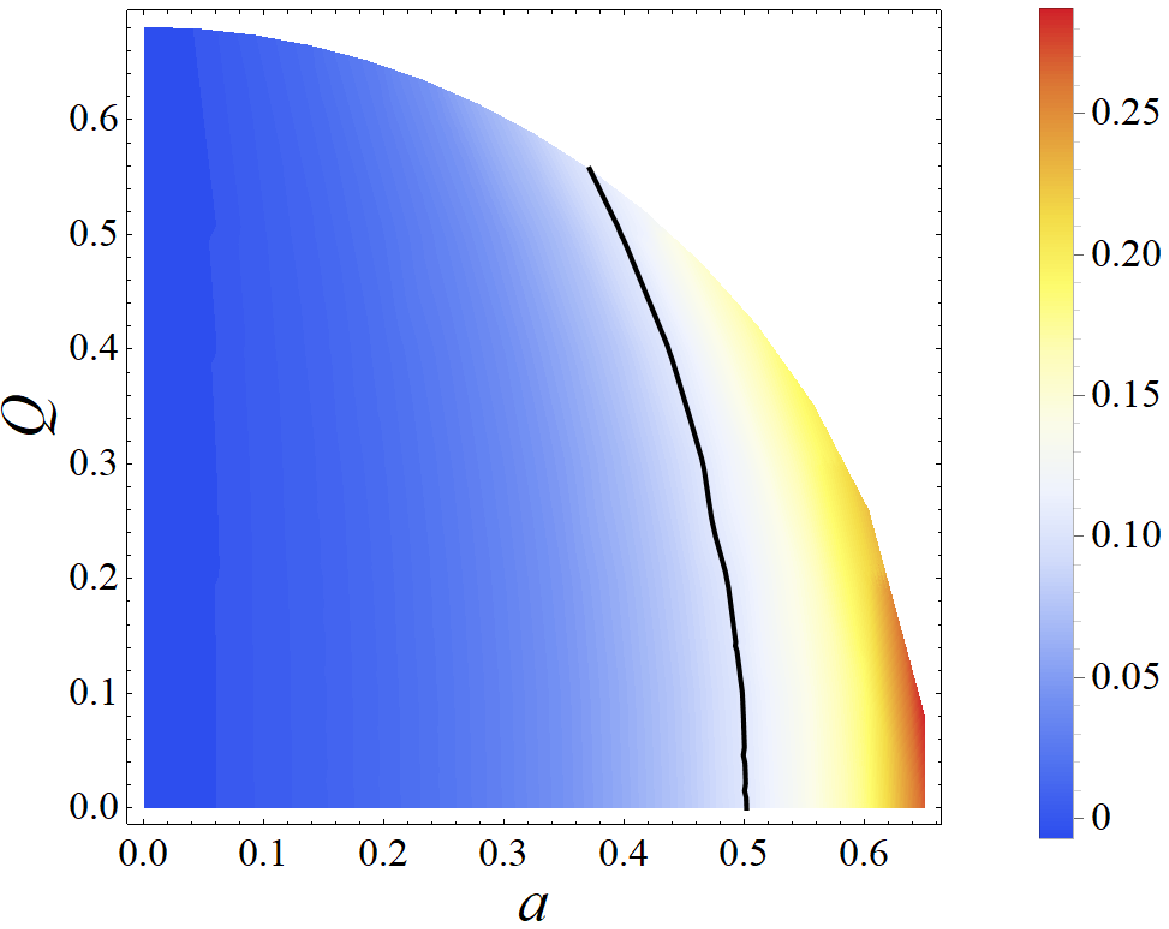}
	\end{tabular}
	\caption{The deviation from circularity $\Delta C$ as a function of $a$ and $Q$ for $\ell=0.0$ (left panel) and $\ell=0.50$ (right panel). The black solid line corresponds to  $\Delta C=0.10$.  }\label{M871}
\end{figure}
We calculate the deviation from circularity $\Delta C$ for the charged rotating Hayward black hole over the entire parameter space for $\theta_0=\pi/2$. In Figs.~\ref{M87} and \ref{M871}, we have shown  $\Delta C$, respectively, as a function of ($a,l$) and ($a,Q$). It is evident from the figures that the measured deviation $\Delta C\leq 0.1$ constrained the black hole parameter space, i.e., the rotating charged Hayward black hole for particular values of parameters can mimic the asymmetry in the observed shadow of the M87* black hole.

\section{Gravitational deflection of light by the charged rotating Hayward black hole}\label{sec6}

\begin{figure}
	\includegraphics[scale=0.7]{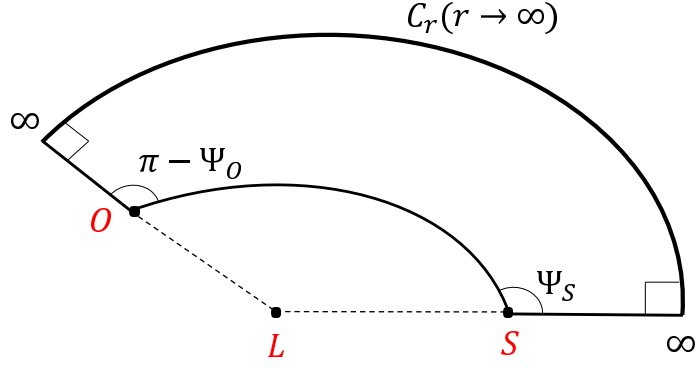}
\caption{Schematic figure for the quadrilateral ${}_O^{\infty}\Box_{S}^{\infty}$ embedded in the curved space. }\label{lensing1}
\end{figure}

The deflection angle for the rotating axisymmetric spacetime at the equatorial plane can be written in terms of the angle made by light rays tangent to the radial direction at the observer ($O$) and source ($S$) positions $\Psi_O$ and $\Psi_S$, respectively, and the angular coordinate separation of the observer and source $\Phi_{OS}$ as \cite{Ono:2017pie}
\begin{equation}
\alpha_D=\Psi_O-\Psi_S+\Phi_{OS}.
\end{equation}
Here, $\Phi_{OS}=\Phi_O-\Phi_S$, where $\Phi_O$ and $\Phi_S$ are, respectively, the angular coordinates of the observer and the source. We consider a quadrilateral ${}_O^{\infty}\Box_{S}^{\infty}$ of spatial light ray curve from the observer and source, and a circular arc segment $C_r$ of coordinate radius $r_C$ $(r_C\to\infty)$ (cf. Fig.~\ref{lensing1}). 
We assume that the source and observer are located at a finite distance from the lens object (black hole), light rays propagating from the source to the observer get deflected due to the gravitational field of lens object. Indeed, this deflection angle explicitly depends on the impact parameter of light, and for large impact parameter $\xi>> \xi_s$, the deflection angle is small. However, as the impact parameter approaches the critical value, the deflection angle gets larger and larger and become unboundedly large for $\xi=\xi_s$ \cite{Virbhadra:1999nm}. These light rays can be described as the spatial curves on a 3-dimensional Riemannian manifold $^{(3)}\mathcal{M}$ described by the optical metric \cite{Gibbons:2008rj}. One can solve the metric (\ref{rotbhtr}) for null geodesics $ds^2=0$ to get
 \begin{equation}
dt= \pm\sqrt{\gamma_{ij}dx^i dx^j}+N_i dx^i,
\end{equation}
where $\gamma_{ij}$ can be identified as the optical metric and $N^i$ as the one-form, respectively, defined by
\begin{eqnarray}
\gamma_{ij}dx^i dx^j&=&\frac{\Sigma^2}{\Delta(\Delta-a^2\sin^2\theta)}dr^2+\frac{\Sigma^2}{\Delta-a^2\sin^2\theta}d\theta^2\nonumber\\ 
&+& \left(r^2+a^2+\frac{2m(r)ra^2\sin^2\theta}{\Delta-a^2\sin^2\theta}\right)\frac{\Sigma\sin^2\theta}{(\Delta-a^2\sin^2\theta)} d\phi^2,\\
N_idx^i&=&-\frac{2m(r)ar\sin^2\theta}{\Delta-a^2\sin^2\theta}d\phi.\label{metric3}
\end{eqnarray}
In the weak field limit, we can define the mass function from Eq.~(\ref{CHmetric}) up to the leading order contributions  
$$
m(r)=M-\frac{Q^2}{2r}-\frac{2M^2l^2}{r^3}.
$$ 
Using the optical metric and Gauss-Bonnet theorem \cite{Carmo, Gibbons:2008rj,Ishihara:2016vdc, Ishihara:2016sfv}, we define the deflection angle for light in terms of the Gaussian curvature $K$ of the surface of light propagation and the geodesics curvature $k_g$ of light curves, which yields \cite{Ono:2017pie}
\begin{equation}
\alpha_D=-\int\int_{{}_O^{\infty}\Box_{S}^{\infty}} K dS+\int_{S}^{O} k_g dl,\label{deflectionangle}
\end{equation}
where $dS$ and $dl$ are, respectively, the infinitesimal area element of the surface and line element along the curve.

 The domain of integration in Eq.~(\ref{deflectionangle}) is a quadrilateral ${}_O^{\infty}\Box_{S}^{\infty}$ in the curved space defined by $\gamma_{ij}$, as shown in Fig.~\ref{lensing1}. The line element along the curve can be identified with the affine parameter for the light rays \cite{Ono:2017pie}. For instance, we study the light propagation in the equatorial plane by setting $\theta=\pi/2$; this allows us to define the Gaussian curvature of the two-dimensional surface as \cite{Werner:2012rc}
\begin{eqnarray}
K&=&\frac{{}^{3}R_{r\phi r\phi}}{\gamma},\nonumber\\
&=&\frac{1}{\sqrt{\gamma}}\left(\frac{\partial}{\partial \phi}\left(\frac{\sqrt{\gamma}}{\gamma_{rr}}{}^{(3)}\Gamma^{\phi}_{rr}\right) - \frac{\partial}{\partial r}\left(\frac{\sqrt{\gamma}}{\gamma_{rr}}{}^{(3)}\Gamma^{\phi}_{r\phi}\right)\right),
\end{eqnarray}
where $\gamma$ is the determinant of a $(2\times 2)$ metric defined at the equatorial plane. $K$ is computed as 
\begin{eqnarray}
K&=&\left(\frac{3Q^2}{r^4}+\frac{8a^2Q^2}{r^6} \right)-\left(\frac{2}{r^3}+\frac{6a^2}{r^5}+\frac{6Q^2}{r^5} \right)M+\left(\frac{3}{r^4}-\frac{6a^2}{r^6} -\frac{20Q^2}{r^6} +\frac{40l^2}{r^6}\right)M^2\nonumber\\
&&+\mathcal{O}\left( \frac{a^2l^2M^2}{r^8},\frac{Q^2l^2M^2}{r^8},\frac{a^2Q^2M^2}{r^8},\frac{M^3}{r^5}\right).
\end{eqnarray}
We have used the weak-field approximation and calculated only the leading order contributing terms.
The integral of Gaussian curvature over the closed quadrilateral reads \cite{Ono:2017pie}
\begin{equation}
\int\int_{{}_O^{\infty}\Box_{S}^{\infty}} K dS= \int_{\phi_S}^{\phi_O}\int_{\infty}^{r_0} K \sqrt{\gamma}dr d\phi,\label{Gaussian}
\end{equation}
where $r_0$ is the closed distance  to the black hole. Using Eqs.~(\ref{phiuch}) and (\ref{r}) and introducing $u=1/r$, we find that the light orbit equation reads 
\begin{equation}
\left(\frac{du}{d\phi}\right)^2=F(u),\label{orbit}
\end{equation}
with 
\begin{equation} F(u)=\frac{u^4\Delta^2\left(\left((1+a^2u^2)-ab\right)^2-\Delta u^4(a-b)^2 \right)}{\Big(a\left((1+a^2u^2)-ab\right)-\Delta u^4(a-b)\Big)^2},
\end{equation}
where $b\equiv \xi$ is the impact parameter. In the weak field limit, Eq. (\ref{orbit}) admits the solution $u=(\sin\phi)/b +\mathcal{O}(M,M^2)$ \cite{Ono:2017pie}, and we can rewrite Eq.~(\ref{Gaussian}) as follows
\begin{equation}
\int\int_{{}_O^{\infty}\Box_{S}^{\infty}} K dS= \int_{\phi_S}^{\phi_O}\int_{0}^{\frac{\sin\phi}{b}}-\frac{K\sqrt{\gamma}}{u^2}du d\phi,
\end{equation}
which for metric (\ref{metric3}) reads as
\begin{eqnarray}
\int\int K dS&=&\frac{2M}{b}\left(\sqrt{1-b^2u_o^2}+\sqrt{1-b^2u_s^2}\right) +\frac{2Ma^2}{3b^3}\left((2+b^2u_o^2)\sqrt{1-b^2u_o^2}+(2+b^2u_s^2)\sqrt{1-b^2u_s^2} \right)\nonumber\\
&-& \frac{MQ^2}{3b^3}\left((16+b^2u_o^2)\sqrt{1-b^2u_o^2} +(16+b^2u_s^2)\sqrt{1-b^2u_s^2}  \right)\nonumber\\
&-&\frac{11a^2Q^2M}{25b^5}\left((3b^4u_o^4+4b^2u_o^2+8)\sqrt{1-b^2u_o^2}+(3b^4u_s^4+4b^2u_s^2+8)\sqrt{1-b^2u_s^2} \right)\nonumber\\
&-&\left(\frac{3Q^2}{4b} +\frac{M^2}{4b}\right)\left(u_o\sqrt{1-b^2u_o^2}+u_s\sqrt{1-b^2u_s^2}\right)\nonumber\\
&-&\left(\cos^{-1}bu_o+\cos^{-1}bu_s\right)\left(\frac{3Q^2}{4b^2} +\frac{3a^2Q^2}{4b^4} -\frac{15M^2}{4b^2}-\frac{9M^2a^2}{4b^4}+\frac{15M^2l^2}{4b^4}-\frac{27M^2Q^2}{64b^4}\right)\nonumber\\
&+&\left(-\frac{a^2Q^2}{4b^3}+\frac{3M^2a^2}{4b^3}-\frac{5M^2l^2}{4b^3}\right)\left(u_o(3+2b^2u_o^2)\sqrt{1-b^2u_o^2}+u_s(3+2b^2u_s^2)\sqrt{1-b^2u_s^2}\right)\nonumber\\
&+&\frac{M^2Q^2}{64b^3}\left((411+146b^2u_o^2)u_o\sqrt{1-b^2u_o^2}+(411+146b^2u_s^2)u_s\sqrt{1-b^2u_s^2} \right)\nonumber\\
&+&\mathcal{O}\left( \frac{a^2l^2M^2}{b^6},\frac{Q^2l^2M^2}{b^6},\frac{a^2Q^2M^2}{b^6},\frac{M^3}{b^3}\right).\label{Gaussian1}
\end{eqnarray}
Here, $u_o$ and $u_s$ are, respectively, the reciprocal of the observer and source distances from the black hole, and we have used $\cos\phi_o=-\sqrt{1-b^2u_o^2}, \cos\phi_s=\sqrt{1-b^2u_s^2}$. The geodesic curvature of light curve with the optical metric in the manifold $^{(3)}\mathcal{M}$ can be described as \cite{Ono:2017pie}
\begin{equation}
k_g=-\frac{1}{\sqrt{\gamma\gamma^{\theta\theta}}}N_{\phi,r},
\end{equation}
which clearly vanishes for a nonrotating black hole spacetime and makes a finite and crucial contribution to the deflection angle around a rotating black hole. The geodesic curvature for metric (\ref{metric3}) is given by
\begin{equation}
k_g=-\frac{2aM}{r^3}-\frac{2aM^2}{r^4} +\frac{2aQ^2}{r^4}+\frac{16aM^2l^2}{r^6}-\frac{60aQ^2M^2}{r^6}+\mathcal{O}\left(\frac{M^3a}{r^5},\frac{aM^2Q^2l^2}{r^8}\right).
\end{equation}
The contribution from the geodesic curvature is in the form of a path integral along the light curve from the source to the observer. Considering a coordinate system centered at the lens position, we can approximate the light curve with $r=b/{\cos\vartheta}$ and $l=b\tan\vartheta$ \cite{Ono:2017pie}. Then, the path integral of geodesic curvature reads 
\begin{eqnarray}
\int_S^O k_g dl &=& \int_S^O \Big(-\frac{2Ma}{b^2}\cos\vartheta-\frac{2M^2a}{b^3}\cos^2\vartheta+\frac{2aQ^2}{b^3}\cos^2\vartheta+\frac{16 aM^2l^2}{b^5}\cos^4\vartheta-\frac{60aQ^2M^2}{b^5}\cos^4\vartheta 
\nonumber\\
& & + \mathcal{O}\left(\frac{M^3a}{b^4}\right)\Big)d\vartheta.\nonumber\\
&=&-\frac{2Ma}{b^2}\left(\sqrt{1-b^2u_o^2}+\sqrt{1-b^2u_s^2}\right)\nonumber\\
&&+\left(\frac{aQ^2}{b^2}-\frac{M^2a}{b^2}\right)\left(u_o\sqrt{1-b^2u_o^2}+u_s\sqrt{1-b^2u_s^2}\right)\nonumber\\
&&+\left(\frac{aQ^2}{b^3}-\frac{M^2a}{b^3}-\frac{45aQ^2M^2}{2b^5}+\frac{6al^2M^2}{b^5}\right)\left( \cos^{-1}bu_s+ \cos^{-1}bu_o\right)\nonumber\\
&&-\frac{15aQ^2M^2}{2b^4}\left(u_o(3+2b^2u_o^2)\sqrt{1-b^2u_o^2}+u_s(3+2b^2u_s^2)\sqrt{1-b^2u_s^2} \right)\nonumber\\
&&+ \frac{2al^2M^2}{b^4}\left(u_o(3+2b^2u_o^2)\sqrt{1-b^2u_o^2}+u_s(3+2b^2u_s^2)\sqrt{1-b^2u_s^2} \right) +\mathcal{O}\left(\frac{M^3a}{b^4}\right),
\label{geodesiccurvature}
\end{eqnarray}   
where we have considered the prograde motion of photons ($dl>0$), for retrograde motion ($dl<0$) we will get an extra -ve sign in Eq.~(\ref{geodesiccurvature}). 
Inserting Eqs.~(\ref{Gaussian1}) and (\ref{geodesiccurvature}) into Eq.~(\ref{deflectionangle}), we can compute the deflection angle for light at the equatorial plane for the finite-distance case, which gives a lengthy expression. In the asymptotically far distance limit, $u_o\to 0$ and $u_s\to 0$, the deflection angle yields
\begin{eqnarray}
\alpha_D&=&\left.\alpha_D\right|_{\text{Kerr}}-\left(\frac{3\pi Q^2}{4b^2}-\frac{a\pi Q^2}{b^3}+\frac{3\pi a^2 Q^2}{4b^4} \right)-\left(\frac{32Q^2}{3b^3}+\frac{176a^2Q^2}{25b^5} \right)M \nonumber\\
&&+ \left(\frac{27\pi Q^2}{64 b^4}-\frac{15\pi l^2}{4b^4}+\frac{6\pi al^2}{b^5}-\frac{45\pi aQ^2}{2b^5}\right)M^2+\mathcal{O}\left(\frac{Q^2a^2M^2}{b^6},\frac{Q^2l^2M^2}{b^6},\frac{M^3}{b^3} \right),~\label{deflection}
\end{eqnarray}
where $\left.\alpha_D\right|_{\text{Kerr}}$ stands for the Kerr deflection angle \cite{Ono:2017pie}
\begin{equation}
\left.\alpha_D\right|_{\text{Kerr}}=\left(\frac{4}{b} -\frac{4a}{b^2}+\frac{8a^2}{3b^3}\right)M+\left( \frac{15\pi}{4b^2}-\frac{a\pi}{b^3}+\frac{9\pi a^2}{4b^4}\right)M^2 +\mathcal{O}\left(\frac{M^3}{b^3}, \frac{M^4}{b^4}\right).
\end{equation}
The deflection angle for a Kerr-Newman black hole ($\ell=0$) can be determined from Eq.~(\ref{deflection}) as 
\begin{eqnarray}
\left.\alpha_D\right|_{\text{KN}}&=&\left(\frac{a\pi Q^2}{b^3}-\frac{3\pi Q^2}{4b^2}-\frac{3\pi a^2 Q^2}{4b^4} \right) + \left(\frac{4}{b} -\frac{4a}{b^2}+\frac{8a^2}{3b^3}+ \frac{32Q^2}{3b^3}+\frac{176a^2Q^2}{25b^5}\right)M\nonumber\\
&&+\left( \frac{15\pi}{4b^2}-\frac{a\pi}{b^3}+\frac{9\pi a^2}{4b^4}+\frac{27\pi Q^2}{64 b^4}-\frac{45\pi aQ^2}{2b^5}\right)M^2 +\mathcal{O}\left(\frac{M^3}{b^3}, \frac{M^4}{b^4}\right).
\end{eqnarray}
In addition, the deflection angle of light for a nonrotating ($a=0$) charged Hayward black hole can be computed from Eq.~(\ref{deflection}) to get
\begin{equation}
\alpha_D=-\frac{3\pi Q^2}{4b^2}+\left(\frac{4}{b}-\frac{32 Q^2}{3b^3}\right)M+\left(\frac{15\pi}{4b^2}-\frac{15\pi l^2}{4b^4}+\frac{27\pi Q^2}{64b^4}+\frac{65\pi l^2Q^2}{16b^6}  \right)M^2+\mathcal{O}\left(\frac{M^3}{b^3}\right),
\end{equation}
which further in the limiting case of $l=0, Q=0$ naturally gives the value for the Schwarzschild black hole \cite{Virbhadra:1999nm} as
\begin{equation}
\left.\alpha_D\right|_{\text{Schw}}=\frac{4M}{b}+\frac{15\pi M^2}{4b^2}+\mathcal{O}\left(\frac{M^3}{b^3} \right).
\end{equation}

\begin{table}[]
	\begin{tabular}{|l|c|c|l|c|}
		\hline
		$Q/M $  & $a/M=0.0$    & $a/M=0.2$     & $a/M=0.4$     &$ a/M=0.6 $    \\
		\hline
		0.0&	$2.42\times 10^{-8}$ & 0.16446 & 0.328875 & 0.493247 \\ \hline
		0.1&	0.00486252 & 0.169321 & 0.333735 & 0.498106 \\  \hline
		0.2&    0.01945 & 0.183905 & 0.348315 & 0.512682 \\  \hline
		0.3&    0.0437625 & 0.208211 & 0.372615 & 0.536975 \\  \hline
		0.4&    0.0777999 & 0.242239 & 0.406634 & 0.570985 \\  \hline
		0.5&   	0.121562 & 0.28599 & 0.450373 & 0.614713 \\  \hline
		0.6&   	0.17505 & 0.339463 & 0.503832 & 0.668158 \\  \hline
	\end{tabular}
	\caption{The corrections in the deflection angle $\delta\alpha_D =\left.\alpha_D\right|_{\text{Schw}}-\alpha_D$ for Sgr A* with $b=10^3M$, $Q=0.30M$ and varying $\ell/M$ and $a/M$, $\delta\alpha_D$ is in units of $as$.}\label{table4}
\end{table}

\begin{table}
	\begin{tabular}{|l|c|c|l|c|}
		\hline
	$	Q/M $  & $a/M=0.0$   & $a/M=0.2$   &$ a/M=0.4  $ &$ a/M=0.6$   \\ \hline
		0.1 & 4.86249 & 4.8612  & 4.85991 & 4.85862 \\ \hline
		0.2 & 19.45   & 19.4448 & 19.4397 & 19.4345 \\ \hline
		0.3 & 43.7624 & 43.7508 & 43.7392 & 43.7276 \\ \hline
		0.4 & 77.7999 & 77.7793 & 77.7586 & 77.738  \\ \hline
		0.5 & 121.562 & 121.53  & 121.498 & 121.466 \\ \hline
		0.6 & 175.05  & 175.00  & 174.957 & 174.91  \\ \hline
	\end{tabular}
\caption{The corrections in the deflection angle $\delta\alpha_D =\left.\alpha_D\right|_{\text{Kerr}}-\alpha_D$ for Sgr A* with $b=10^3M$ and $l=0.10M$, $\delta\alpha_D$ is in units of $mas$. }\label{table2}
\end{table}

\begin{table}[]
	\begin{tabular}{|l|c|c|l|c|}
		\hline
		$\ell/M $  & $a/M=0.0$    & $a/M=0.3$     & $a/M=0.4$     &$ a/M=0.6 $    \\ \hline
		0.1 & 0.024203 & 0.0241914 & 0.0241876 & 0.0241797 \\ \hline
		0.2 & 0.09681  & 0.0967652 & 0.0967498 & 0.0967186 \\ \hline
		0.3 & 0.217826 & 0.217722  & 0.217687  & 0.217617  \\ \hline
		0.4 & 0.387247 & 0.387061  & 0.386999  & 0.386875  \\ \hline
		0.5 & 0.605073 & 0.604782  & 0.604686  & -         \\ \hline
		0.6 & 0.871305 & 0.870887  & 0.870747  & -         \\ \hline
	\end{tabular}
\caption{The corrections in the deflection angle $\delta\alpha_D =\left.\alpha_D\right|_{\text{KN}}-\alpha_D$ for Sgr A* with $b=10^3M$, $Q=0.30M$ and allowed values of $\ell/M$ and $a/M$, $\delta\alpha_D$ is in units of $\mu as$.}\label{table3}
\end{table}

\begin{figure}
	\includegraphics[scale=0.9]{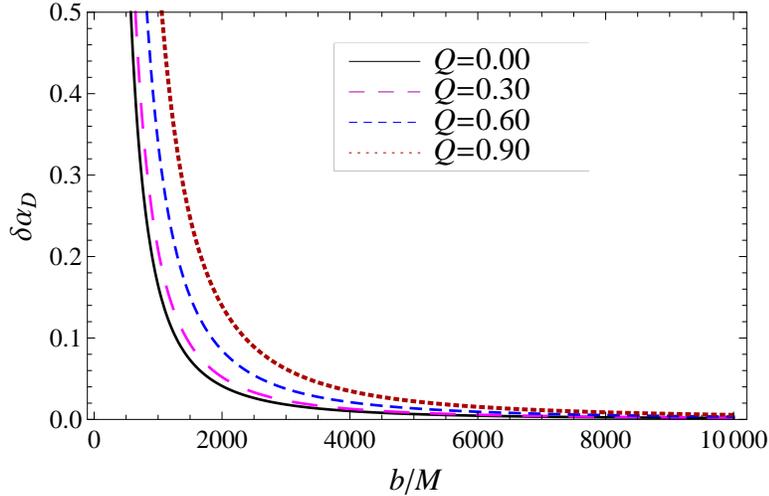}
	\caption{The correction in the deflection angle $\delta\alpha_D =\left.\alpha_D\right|_{\text{Schw}}-\alpha_D$ variation with $b$ for  $a=0.2M$ and $l=0.2M$.}\label{DefAng}
\end{figure}
In order to discuss the possible astronomical applications, we model the Sgr A* ($M=4.6\times 10^6 M_{\odot}$) as a charged rotating Hayward black hole and calculate the deflection angle of light for varying parameters $a$, $Q$ and $\ell$. The corrections in the deflection angle for the charged rotating Hayward black hole from the Schwarzschild, Kerr and Kerr-Newman black holes are determined,  i.e, $\delta\alpha_D=\left.\alpha_D\right|_{\text{Schw}}-\alpha_D$ in Table \ref{table4}, $\delta\alpha_D=\left.\alpha_D\right|_{\text{Kerr}}-\alpha_D$ in Table \ref{table2}, and $\delta\alpha_D=\left.\alpha_D\right|_{\text{KN}}-\alpha_D$ in Table \ref{table3}. For fixed values of the black hole parameters and the impact parameter, the charged rotating Hayward black hole caused a smaller deflection angle than that for the Schwarzschild, Kerr and Kerr-Newman black holes. The order-of-magnitude of corrections made by the nonzero $Q$ and $\ell$ are $as$ and $\mu as$, respectively. In Fig.~\ref{DefAng}, we have shown  $\delta\alpha_D=\left.\alpha_D\right|_{\text{Schw}}-\alpha_D$ with varying $b$ for different values of the black hole parameter. As expected, the effect of $Q$ is prominent only for a small impact parameter.

\section{Conclusion}\label{sec7}

The solution of Einstein's field equations for a source satisfying the generic energy conditions exhibit both past and future singularities encompassed by the event horizon \cite{Hawking:1969sw,Clarke:1975ph}. The singularity pathology in general relativity, which indicates the breakdown of the classical theory and requires modifications at high energies, motivated physicists to develop the idea of regular spacetimes inside black holes. As expected the resulted energy-momentum tensor should violate some of the energy conditions, and though the solutions are deprived of central curvature singularity, horizons may still present. Hayward's black hole solution is one such example, whose global structure is very similar to a singular black hole, namely, to the Reissner-Nordstr\"{o}m black hole except that now $r=0$ is a regular point \cite{DeLorenzo:2014pta}.

In this paper, we have studied the charged rotating Hayward black hole, which realizes the Reissner-Nordstr\"{o}m, Kerr and Kerr-Newman black hole solutions as various limiting cases. The black hole solution interpolates between a de-Sitter core at small $r$ and a Kerr-Newman black hole at large $r$. The obtained black hole solution has up to two horizons,  provided that the  parameters are properly chosen. Furthermore, there always exist extremal values of the parameters, for which both horizons coincide. The allowed parameter space ($a, \ell$) for the existence of black hole horizons becomes more and more compact with increasing $Q$. We found that the black hole is supported by a physically reasonable source whose components are well defined, bounded from above and fall appropriately at large distances. For the rotating solution, the weak energy condition may be violated and the violation becomes stronger near the central region. The nonzero value of charge $Q$ has a profound impact on this violation, as the region of violation ($r\leq r_c$) increases with $Q$. It is shown that for rotating charged Hayward black holes, the region of violation is always inside the Cauchy horizon, i.e., $r_c<r_-$. 

Employing the spacetime isometries, we determined the corresponding effective mass $M_{eff}$ and angular momentum $J_{eff}$. These entities calculated within a finite $r$ are found to be lower than their asymptotic values. Moreover, at a fixed radial coordinate $r$, the values of  $M_{eff}$ and $J_{eff}$ for the charged rotating Hayward black hole are smaller than those for the Kerr-Newman black holes. Emphasizing on the observational signatures that a charged regular black hole can have, gravitational lensing and the resulting black hole shadow are discussed. For lensing, we considered that the source and the observer are at finite distances from the black hole. We found that for fixed values of parameters, the deflection angle of light for a rotating charged Hayward black hole is smaller than those for the Kerr or Kerr-Newman black holes. In particular, the correction made by the nonzero $\ell$ to the deflection angle in the weak-field limit is of the micro-arcsecond order. Shadows for various values of parameters are constructed and it is found that the presence of $Q$ and $\ell$ made noticeable changes in the shadow shape and size. In particular, the shadow gets smaller and more distorted comparing to that for the Kerr black hole with increasing $Q$ or $\ell$. With the aid of the recent M87* black hole shadow observations, we modeled the rotating charged Hayward black hole as M87* and put constraints on the parameters of the solution.

\begin{acknowledgements}  S.G.G. would like to thank DST INDO-SA bilateral project DST/INT/South Africa/P-06/2016 and also IUCAA, Pune for the hospitality while this work was being done. R.K. would like to thank UGC for providing SRF. A.W. thanks the staff at ZJUT for its hospitality, at which part of the work was done and also thanks the National Natural Science Foundation of China for partial support with Grant No. 11375153 and No. 11675145.
\end{acknowledgements}

\noindent
\end{document}